\documentclass[12pt]{article}
\usepackage{natbib}
 \bibpunct{(}{)}{;}{a}{,}{,}
\usepackage{graphicx}
\usepackage{amsmath,amssymb,amsthm}
\usepackage{bm}
\usepackage{rotating}
\usepackage{rotating}
\usepackage{multicol}
\usepackage{dsfont}
\usepackage{titlesec}
\usepackage{latexsym}
\usepackage{changepage}
\usepackage{multirow}
\usepackage{booktabs}
\usepackage{pbox}
\usepackage[affil-it]{authblk}
 \usepackage[latin1]{inputenc}
\usepackage[titletoc]{appendix}
\usepackage{float}
\usepackage{caption}
\usepackage[linktoc=page,colorlinks,linkcolor=blue,citecolor=blue,urlcolor=blue]{hyperref}
\usepackage{fancyhdr}
\usepackage{setspace}
\usepackage[top=1in, bottom=1in, left=1in, right=1in]{geometry}
\usepackage{enumitem}
\usepackage[normalem]{ulem}
\usepackage[table]{xcolor}
\usepackage{color}

\newcommand{\bv}{\begin{array}}
\newcommand{\ev}{\end{array}}
\newcommand{\bit}{\begin{itemize}}
\newcommand{\eit}{\end{itemize}}
\newcommand{\ben}{\begin{enumerate}}
\newcommand{\een}{\end{enumerate}}
\newcommand{\beq}{\begin{equation}}
\newcommand{\eeq}{\end{equation}}
\newcommand{\bvq}{\begin{eqnarray}}
\newcommand{\evq}{\end{eqnarray}}

\newcommand{\E}{\mathbb{E}}

\allowdisplaybreaks


\newcounter{parentnumber}

\theoremstyle{plain}

\newtheorem{assumption}{Assumption}

\newcommand{\vZ}{\textbf{Z}}
\newcommand{\vY}{\textbf{Y}}
\newcommand{\vX}{\textbf{X}}
\newcommand{\vS}{\textbf{S}}
\newcommand{\bs}{\textbf{s}}

\newcommand{\vG}{\textbf{G}}

\newcommand{\ind}{\perp\!\!\!\perp}


\newcommand\SOTwo{SO$_2$ }
\newcommand\PMTwo{PM$_{2.5}$ }
\newcommand\numbeneficiaries{21,577,552 } 
\newcommand\numplants{472 } 
\newcommand\numscrubbers{106 } 
\newcommand\numkeyassoc{278 } 
\newcommand\numkeyassocscrubbed{35 } 
\newcommand\numzips{25,553 }
\newcommand\numzipswithkeytreated{2,753 } 

\newcommand{\jstar}{j_{(i)}^*}

\begin{document}
\pagestyle{empty}

\title{Bipartite Interference and Air Pollution Transport: Estimating Health Effects of Power Plant Interventions}
  \author[1]{Corwin Zigler}
 \author[1]{Vera Liu}
 \author[2]{Fabrizia Mealli}
 \author[3]{Laura Forastiere}

 \affil[1]{University of Texas at Austin}
 \affil[2]{University of Florence}
    \affil[3]{Yale University}

\date{}


\maketitle
\begin{abstract}
Evaluating air quality interventions is confronted with the challenge of interference since interventions at a particular pollution source likely impact air quality and health at distant locations and air quality and health at any given location are likely impacted by interventions at many sources. The structure of interference in this context is dictated by complex atmospheric processes governing how pollution emitted from a particular source is transformed and transported across space, and can be cast with a bipartite structure reflecting the two distinct types of units: 1) {\it interventional units} on which treatments are applied or withheld to change pollution emissions; and 2) {\it outcome units} on which outcomes of primary interest are measured.  We propose new estimands for bipartite causal inference with interference that construe two components of treatment: a ``key-associated'' (or ``individual'') treatment and an ``upwind'' (or ``neighborhood'') treatment. Estimation is carried out using a semi-parametric adjustment approach based on joint propensity scores.  A reduced-complexity atmospheric model is deployed to characterize the structure of the interference network by modeling the movement of air parcels through time and space. The new methods are deployed to evaluate the effectiveness of installing flue-gas desulfurization scrubbers on \numplants coal-burning power plants (the interventional units) in reducing Medicare hospitalizations among \numbeneficiaries Medicare beneficiaries residing across \numzips ZIP codes in the United States (the outcome units). 
\vskip 1ex
\noindent\textbf{Keywords:} Causal inference;  Network interference; Generalized propensity scores; Air pollution; Power plants.
\end{abstract}


\pagestyle{fancy}
\setcounter{page}{1}


\section{Introduction}
Evaluating public-health interventions is increasingly challenged by inherent interconnectedness of observational units, often cast as a network with observational units as nodes and connections between them as edges.   Causal inference in such settings may confront {\it interference}, which arises when outcomes for some units depend in part on treatments applied to other units.  The most typical examples include vaccine interventions with effects propagating across infection networks of individuals who come into contact with one another and informational interventions on individuals connected through their social network. 

We consider the evaluation of public-health interventions to reduce harmful pollution emissions from power plants.  Interference in this case arises due to phenomena known as {\it pollution transport and chemistry}; chemical compounds such as sulfur dioxide (\SOTwo) emitted from a power plant smokestack are transported through the atmosphere and react chemically with atmospheric co-constituents. The primary end product of atmospheric \SOTwo is sulfate (SO$_4^{2-}$), which condenses quickly and contributes to increased fine particulate air pollution (\PMTwo), a pollutant at the center of many regulatory policies owing to its link to myriad health end points \citep{pope_iii_fine-particulate_2009, dominici_particulate_2014}. Therefore, an intervention employed at one power plant will likely affect health outcomes in the locations where chemical compounds are transported, and the health outcomes at a given location are dictated in part by actions taken at many power plants. 

The motivating power plant setting introduces three key methodological challenges at the center of this work that expands the statistical literature on causal inference with interference. First, the common cases of clustered or stratified interference \citep{hudgens_toward_2008, perez-heydrich_assessing_2014, tchetgen_causal_2012, papadogeorgou_causal_2019} are not appropriate for the power plant setting, placing this work in vein of  recent efforts to consider more general structures of interference \citep{van_der_laan_causal_2014, sofrygin_semi-parametric_2017, forastiere_estimating_2022, forastiere_identification_2021, ogburn_causal_2020, tchetgen_tchetgen_auto-g-computation_2021, savje_average_2021}.  Second, in contrast to the often-considered setting where interference arises due to unit-to-unit outcome dependencies (e.g., as in an infectious disease), interference in this work is dictated by complex physical/mechanistic process - here, the transport of chemical air pollution - for which we deploy a reduced-complexity atmospheric dispersion model to characterize the complex network giving rise to the interference.  This feature has points of contact with a budding literature on so-called ``spatial interference'' \citep{verbitsky-savitz_causal_2012,  giffin_generalized_2020, aronow_design-based_2020, zirkle_addressing_2021}.  In particular, \citet{aronow_design-based_2020} considers a setting most similar to that considered here, specifying generic nonparametric and parametrically-smoothed functions for ``ambient effects'' emanating from treatment points, with focus tailored to estimands and estimation for design-based inference in randomized experiments. A key distinction of our work (in addition to those mentioned below) is the characterization of structural network interference with the combination of geography and atmospheric conditions in a manner that goes beyond just spatial proximity.   Finally, and most notably, we consider the setting of {\it bipartite causal inference with interference} \citep{zigler_bipartite_2021}, where the network of observational units consists of two distinct types: {\it interventional units}, on which treatments are applied or withheld, and {\it outcome units} on which outcomes of interest are defined and measured.  Explication of the bipartite setting has only recently appeared, with consideration beyond air pollution including most notably online marketplace experiments \citep{pouget-abadie_variance_2019, doudchenko_causal_2020, harshaw_design_2021}.  All of the above challenges are confronted in the context of an observational study without experimental control over the interventions or the structure of interference.  

The case of bipartite causal inference with interference was introduced in  \cite{zigler_bipartite_2021} amid the same motivating power plant problem considered here, wherein inverse probability of treatment weighted (IPTW) estimators hewing closely to existing literature on partial interference \citep{perez-heydrich_assessing_2014, tchetgen_causal_2012, papadogeorgou_causal_2019}  were deployed in the simplified setting where power plants were clustered geographically into non-overlapping groups.  To accommodate a more complex and realistic interference structure reflective of the realities of pollution transport, we continue development from \cite{forastiere_identification_2021} to estimate bipartite versions of direct and indirect (or spillover) effects. The approach allows interference to take place on a structural network and construes the ``treatment'' under investigation in two components: a ``key associated'' treatment specifying a characteristic of the intervention that is specific to an individual location (e.g., whether the power plant having the most impact on that location adopts the intervention), and a ``neighborhood'' or ``upwind'' treatment characterizing treatments among interfering units (e.g., a function of the treatment statuses of power plants located upwind from a location).  Reducing the intervention to these two components helps focus the definition of potential outcomes so that causal estimands and an assignment mechanism can be formalized in the bipartite setting. 
Under an inferential perspective where potential outcomes are viewed as random variables with realized observed values
on the observational units, estimation of causal effects is based on a joint propensity score model for the two treatment components \citep{forastiere_identification_2021}. While the methodology pursued here relies heavily on theoretical results from \cite{forastiere_identification_2021}, the bipartite nature of the problem entails nontrivial differences in formulation of the estimands, assignment mechanism, and implications for different types of confounding.

An important feature of this work is the manner in which we characterize the mechanistic phenomena underlying the structure of interference. To characterize the structure of interference, we deploy a newly-developed reduced-complexity atmospheric model, called HYSPLIT Average Dispersion (HyADS), to model the movement of pollution through space and time \citep{henneman_characterizing_2019}. The characterization of a network that is not based on notions of contacts or adjacency offers potential advantages owing to the interpretability of estimands relying on functions of the interference network, but necessitates careful attention to the definition of useful estimands in the bipartite setting. Combining the atmospheric model for pollution transport with novel methods for bipartite causal inference with interference represents an important advance in the methodology available for evaluating interventions at point sources of air pollution. 

\section{Background and Data for Evaluating Power Plant Interventions}\label{sec:background}

\subsection{Title IV of the Clean Air Act Amendments and Scrubbers on Coal-Fired Power Plants}
Starting with at least Title IV of the 1990 Amendments to the Clean Air Act, air quality management in the US has striven to reduce \SOTwo emissions by ten million tons relative to 1980 levels \citep{chestnut_fresh_2005}.  One motivation for such regulations is the fact that \SOTwo is a known precursor to the atmospheric formation of \PMTwo, which itself has been linked to myriad adverse health outcomes \citep{pope_iii_fine-particulate_2009, dominici_particulate_2014}. Thus, a major focus of such efforts to reduce population pollution exposure is the reduction of \SOTwo (and other) emissions from coal-fired electricity generating power plants, the dominant source of \SOTwo emissions in the US.  

The specific intervention evaluated here is the installation (or not) of flue-gas desulfurization scrubbers (``scrubbers'') on \numplants coal-fired power plants in the United States during 2005, a year of significant regulatory action on power plants.  Such scrubbers are known to reduce emissions of \SOTwo.  We deploy the new methods to estimate network intervention effects of scrubber installation on hospitalization outcomes among \numbeneficiaries adults aged 65 and older enrolled in Medicare and residing across \numzips ZIP codes. 

\subsection{Pollution Transport and HYSPLIT Average Dispersion}\label{sec:hyads}
One key feature of the link between \SOTwo emissions and population health outcomes is the phenomenon of long-range pollution transport, which governs how \SOTwo emissions originating at a specific power plant transport across time and space as \SOTwo reacts chemically to form SO$_4^{2-}$ and ultimately increase ambient \PMTwo to which populations are exposed.  Such transport can render the ambient pollution (and population health) at a particular location susceptible to changes in emissions from power plants located at great distances. Thus, a central task for investigating the impacts of scrubber installation on population health is characterization of which ZIP codes in the US might be affected by scrubber installation at each of the power plants under study.

We use a recently-developed reduced-complexity atmospheric model, called HYSPLIT Average Dispersion (HyADS) to achieve such characterization \citep{henneman_characterizing_2019}.  Briefly, HyADS simulates hundreds of thousands of ``emissions events'' mimicking the release of air mass from the location of each coal power plant smokestack, following each mass forward in time and tracking its movement trajectory (as governed by Lagrangian trajectory mechanics and historical wind field data).  Parcel locations are then linked to geographic locations (e.g., ZIP codes) to generate a metric of the number of times per day each ZIP code is impacted by air originating at each power plant. For this investigation, linked parcel locations for each day are aggregated throughout the entire year of 2005, representing an annual impact of parcels on a given ZIP code location (re-scaled to have maximum value 1). Details on the HyADS approach appear in \cite{henneman_characterizing_2019}, where the approach is shown to have good agreement with state-of-the art chemical transport models for air pollution which cannot generally be employed at the computational scale required for the present investigation.  The end result of the HyADS simulation is output of a ``source-receptor'' matrix with entries between $[0, 1]$ characterizing each power plant's annual influence on each ZIP code. This will form the basis of the network adjacency matrix in Section \ref{sec:interference_map}.  Deriving a network adjacency matrix with information representing a physical/chemical process represents an important point of departure from studies on social networks. 

\subsection{Supporting Data on Power Plants and Zip Codes} In addition to the historical wind fields data underlying the HyADS simulations, data on monthly \SOTwo emissions for each coal-fired power plant operating in the US were obtained from the US Environmental Protection Agency (EPA) Air Markets Program Database, along with information about power plant characteristics, including the dates of any scrubber installations.  HyADS also uses information on the heights of power plant smokestacks, obtained from the US Energy Information Administration.  Medicare health outcomes come from the Center for Medicare and Medicaid Services.  These data were processed into annual counts (and rates) of hospitalizations for each US ZIP code, along with supporting data on person-years at risk for hospitalization as well as basic demographic characteristics of Medicare beneficiaries. For this evaluation, we focus on hospitalizations for Ischemic Heart Disease (IHD) which has been specifically linked to ambient \PMTwo derived from coal combustion in \cite{thurston_ischemic_2016} and \cite{henneman_accountability_2019}. Demographic information on the general population of each ZIP code was obtained from the US Census (year 2000), and county-level smoking rates come from small-area estimated values anchored to the CDC Behavior Risk Factor Surveillance System \citep{dwyer-lindgren_cigarette_2014}. Weather and climatological characteristics for each ZIP code come from the North American Regional Reanalysis \citep{kalnay_ncep/ncar_1996}, and a annual average total mass of \PMTwo (for use in a secondary analysis) is obtained from GEOS-Chem chemical transport model predictions on a grid across the US and linked to the ZIP code level \citep{van_donkelaar_regional_2019}.

\section{Notation and Estimands for the Bipartite Setting}\label{sec:Notation_Estimands}
\subsection{Potential Outcomes On Bipartite Networks}\label{sec:potential_outcomes}
Against the backdrop of the power plant problem and data outlined in Section \ref{sec:background}, we offer here the development of potential outcomes in settings of bipartite interference, as detailed in \cite{zigler_bipartite_2021}. Let $j=1,2,\ldots,J$ index a sample of $J$ observational units, at which a well-defined intervention may or may not occur, with an indicator $S_j=1$ if the $j^{th}$ unit is ``treated'' with the intervention and $S_j=0$ otherwise. Call these observational units {\it interventional units}.  In the motivating power plant example, the interventional units are $J=\numplants$ coal-fired power plants operating in the US in 2005, and $S_j=1$ denotes that the $j^{th}$ plant had a scrubber installed for at least half of the year.  The vector $\vS = (S_1, S_2, \ldots, S_J)$ denotes the vector of treatment assignments to the $J$ interventional units, taking on a specific value $\bs \in \mathcal{S}(J)$, where $\mathcal{S}(J)$ denotes the space of possible such vectors.   Denote covariates measured at the interventional units with $\vX_j^{int}$.  

Let $i=1,2,\ldots,n$ index a second, distinct set of observational units where outcomes of interest are defined and measured.  Call these units {\it outcome units}, and let $Y_i, i =1,2,\ldots,n$ represent an outcome of interest measured at each.  For example, in the power plant investigation, $Y_i$ denotes the number of hospital admissions for ischemeic heart disease (IHD) in 2005 among Medicare beneficiaries residing in each of $n=\numzips{}$ ZIP codes across the US.  Denote covariates measured at the outcome units with $\vX_i^{out}$ for $i=1,2,\ldots,n$.  Settings with observational units, outcomes, and interventions described as above have been referred to as settings of {\it bipartite causal inference} \citep{zigler_bipartite_2021}.  

Note that, without further restrictions or assumptions on the bipartite structure, there is no clear definition of the intervention for the outcome units.  Nonetheless, the general goal will be to estimate causal effects of the intervention, $S$, on the outcome $Y$.  Formalizing such questions can proceed with potential outcomes in the bipartite setting, following in much the same manner as in settings of one level of observational unit.  Let $Y_i(\bs)$ denote the potential outcome that would be observed at outcome unit $i$ under treatment allocation $\bs$, for example, the number of IHD hospitalizations that would occur at the $i^{th}$ ZIP code under a specific allocation of scrubbers to the $J$ power plants.  In full generality, for each outcome unit, the number of potential outcomes $Y_i(\vS)$ correspond to the number of possible allocations in $\mathcal{S}(J)$, for example, $2^J$ possible treatment vectors when $S_j$ is binary and each interventional unit is eligible for treatment.  The key difference owing to the bipartite setting is that $\vS$ is a vector of length $J$, not a vector of length $n$, as would be the case under typical development of potential outcomes with one level of observational unit. Implicit in the above notation is the assumption of consistency or ``no multiple versions of treatment,'' that is, $Y_i(\bs) = Y_i(\bs')$ for all $i$ when $\bs = \bs'$.

In the subsequent, we adopt a model-based perspective for inference \citep{imbens_causal_2015, hernan_causal_2020}, whereby potential outcomes  are regarded as random variables whose observed values are drawn from a  specified model, as is common when estimating causal effects on a fixed set of units corresponding to the whole population of interest \citep{li_bayesian_2022}. Note that this approach can be viewed as the same a superpopulation approach where the sampling reproduces the distribution of outcomes drawn from the model used in the model-based perspective \citep{hernan_causal_2020}.


\subsection{Continuous Interference Mappings for Weighted Directed Networks}\label{sec:interference_map}
Typical formulation of potential outcomes would proceed with the so-called Stable Unit Treatment Value Assumption (SUTVA) clarifying, in part, that there is ``no interference'' between units in the sense that outcomes for a given unit do not depend on treatments applied at other units.  The lack of immediate correspondence between interventional units and outcome units in the bipartite setting precludes an immediate statement of SUTVA.  While fully general development would allow the outcome at any outcome unit to depend on the treatments assigned at all interventional units, there may be information to support constraints on the structure of interference.  These constraints have been previously specified in settings with one type of observational unit with ``interference mappings'' \citep{zigler_bipartite_2021}, ``interference sets'' \citep{liu_inverse_2016}, or ``interference neighborhoods'' \citep{karwa_systematic_2018}, and such information is often coded with a graph specifying a specific network structure where the set of units that interfere with an index unit consists of those with a limited path distance from the index node, typically those that are adjacent ``neighbors'' in the network \citep{forastiere_identification_2021}.  In standard settings, the interference set is specified on a one-mode network, representing interconnections between units. In the bipartite setting, where actions at interventional units can impact outcome units, but not {\it vice versa}, interference mapping should be defined on a different kind of network structure, with two sets of nodes and ties linking nodes belonging to different sets. This structure can be regarded as a {\it bipartite (or two-mode) directed} network.   

Settings where interference arises due to complex exposure patterns invite specification of interference structures that expand beyond discrete notions of interference sets to encode continuous degrees of interference that depend on the propagation or diffusion of exposure across the network. In particular, while interference sets in social networks are typically defined {\it topologically}, we consider settings where interference is more aptly viewed {\it geographically} or {\it physically}, as dictated by a (continuous) underlying process. For example, for interventions applied to spatially-indexed units, the degree of interference between two units may be dictated in part by the geographic distance between them \citep{aronow_design-based_2020, giffin_generalized_2020} or, in the case of the power plant study, the geographic distance and features of the atmospheric processes that transport pollution from sources to populations.  Thus, in the power plant setting, the structure of interconnectedness between interventional and outcome units can be regarded as a {\it bipartite weighted and directed} network. 

For such a bipartite weighted and directed network, we expand the notion of an {\it interference mapping} from \cite{zigler_bipartite_2021}.  Specifically, let $t_i^\top = (t_{i1}, t_{i2}, \ldots, t_{iJ})$ denote outcome-unit specific interference map for the $i^{th}$ outcome unit, with $t_{ij}$ quantifying the weighted connectedness between interventional unit $j$ and the outcomes defined at outcome unit $i$.  The sample interference map can then be defined as $T = (t_1, t_2, \ldots, t_n)^\top$, an $n \times J$ matrix with (i,j) entry indicating the strength of influence of the $j^{th}$ interventional unit on the potential outcome for the $i^{th}$ outcome unit. In the power plant evaluation, the entries of $T$ are generated directly from HyADS simulations, representing the aforementioned source-receptor matrix. Note that this characterization of interference in the power plant setting is based only on wind fields and parcel movement trajectories, and is not affected by scrubber installations, that is, the structure of $T$ does not depend on the treatment allocation $\vS$. 


\subsection{Indexing Potential Outcomes with Treatment Functions on the Bipartite Network}\label{sec:treat_functions}
The bipartite setting's lack of immediate correspondence between a single well-defined treatment for each outcome unit complicates the definition of relevant potential outcomes and causal contrasts above and beyond the difficulty in managing the sheer number of relevant potential outcomes. For example, while the approach of \cite{forastiere_identification_2021} relied on common social-network delineation of the individual (and its treatment) and that individual's first-order neighbors (and their treatments) to define potential outcomes and formulate assumptions about interference, the bipartite case introduces two barriers to this type of formulation.  First, the bipartite case lacks any immediate notion of path distance dictating e.g., a unit's first-order neighbors, so there is no self-evident notion of what constitutes an outcome unit's ``neighborhood.'' Second, the bipartite setting entails no natural  notion of an ``individual'' treatment, since no treatment is directly applied to or withheld from the outcome units. 

We use the structure of the bipartite network, specified with $T$, to outline several relevant notions for how two units could interfere with one another in the bipartite setting.  The most basic notion would be that any $(i,j)$ pair of outcome-interventional units interfere if $t_{ij}>0$. Thus, treatments applied to an outcome unit's interfering interventional units will comprise a notion of ``neighborhood'' treatment.  Additionally, two outcome units $(i,i')$ can interfere if they share an interfering interventional unit, that is, $t_{ij}>0$ and $t_{i'j}>0$ for at least one $j$.  We refer to such $(i,i')$ as having overlapping interference sets.  Analogously, two interventional units $(j,j')$ have overlapping interference sets if $t_{ij}>0$ and $t_{ij'}>0$ for at least one $i$, that is, if they share a interfering outcome unit. These distinctions will be important when formalizing different types of confounding.

To define a notion of ``individual'' treatment, the approach here is to first identify a single interventional unit that might be particularly relevant for each outcome unit, and then follow similar reasoning to that outlined in \cite{forastiere_identification_2021} and \cite{karwa_systematic_2018} for the case of a network or graph defined on one type of observational unit. For each of $i=1,2,\ldots,n$ outcome units, denote the ``key associated'' interventional with $\jstar$ \citep{zigler_bipartite_2021}.  For the present investigation, $\jstar$ will be the power plant that is most influential (as determined by HyADS) for the $i^{th}$ ZIP code (specific definition deferred until Section \ref{sec:analysis}).   Note that, in general, the definition of the key associated interventional unit for the $i^{th}$ outcome unit need not be a function of $T$.  For example, $\jstar$ could be alternatively defined as the power plant that is geographically closest to the $i^{th}$ ZIP code.

Combining each outcome unit's key-associated interventional unit with the above notion of the outcome unit's neighborhood will support definition of functions of treatments on the network to encode assumptions about the interference mechanism in order to: a) reduce the number of potential outcomes required to answer relevant scientific questions and b) define causal estimands that can provide answers to those questions. This has been referred, as in \cite{karwa_systematic_2018}, as specifying an ``exposure model'' to specify how the treatments of those in the interference set impact outcomes of an index unit, and is similar to the ``exposure mapping'' of \cite{aronow_estimating_2017}. 

With definition of the key-associated unit, we define the key-associated treatment variable for each outcome unit, $Z_i=S_{\jstar}$, pertaining to the intervention status of the key-associated interventional unit.  For example, in the power plant investigation, $Z_i$ will denote whether the power plant most influential for the $i^{th}$ ZIP code had a scrubber installed for at least half of 2005. To reflect the additional dependence of potential outcomes on treatments applied at interventional units other than $\jstar$, we introduce another treatment variable characterizing the treatments assigned to other units, in accordance with the information contained in the interference mapping $T$.  Formally, let $g_i(\cdot; T): \{0,1\}^{J-1} \rightarrow \mathcal{G}_i$ be an exposure mapping function that maps, for a given interference mapping, $T$,  the treatments on all $J$ interventional units but the ${\jstar}^{th}$ into a scalar value defined for each outcome unit $i=1,2,\ldots, n$.  Denote with $G_i$ the value of the function $g_i(\mathbf{S}, T)$ for the $i^{th}$ outcome unit.  For example, the power plant investigation will make use of $G_i=\sum_{j\neq \jstar} t_{ij}S_j$ to denote the interference-weighted sum of scrubber installations to interventional units other than the key-associated unit.  While this quantity is closely related to the ``neighborhood treatment" function defined in \cite{forastiere_identification_2021}, we will refer to this function as the ``upwind treatment," corresponding to its (approximate) interpretation in the evaluation of power plant interventions (where the term ``upwind" is used loosely to reflect the information output by HyADS).  A more general term relevant to other settings where interference is due to complex exposure patterns may be ``upstream treatment'', because $G_i$ is usually defined by weighting the treatment vector $\vS$ by the inward link weights $t_i$ of the adjacency matrix $T$. In a typical network setting, $G_i$ might be a function of only the treatments in a first-degree neighborhood of units with a direct link to $i$ (as in \cite{forastiere_identification_2021}).  In the current setting, $G_i$ is a function of the whole treatment vector defined for the interventional units.  In principle, every power plant can have nonzero connection to every ZIP code, with the extent of interference based on HyADS.

The utility of formulating the key-associated treatment, $Z_i$, and the upwind treatment, $G_i$, is that doing so permits a key assumption about potential outcomes that formalizes  interference in the bipartite setting.  Specifically, we adopt the following as an alternative to SUTVA in the case of bipartite causal inference with interference:

\begin{assumption}[Upwind Interference]
\label{ass:SUTNVA}
For a fixed $T$, any two $(\vS, \vS') \in \mathcal{S}(J)$ such that the corresponding $Z_i=Z'_i$ and $G_i = G'_i$ yield the following equality:
\[ \qquad Y_i(\vS) = Y_i(Z_i,G_i) = Y_i(Z'_i, G'_i) = Y_i(\vS')\]
\end{assumption}

In other words, Assumption \ref{ass:SUTNVA} reduces the implications of interference to depend only on the index outcome unit's key-associated treatment and the scalar-valued function of treatments applied to all other interventional units.  In the power plant example, this implies that the IHD hospitalization rate at ZIP code $i$ would be the same under any two allocations of scrubbers to all power plants across the country that produces a specified treatment status of the most influential plant and the same upwind treatment rate.   Since definition and interpretation of the estimands described in the subsequent will rely heavily on Assumption \ref{ass:SUTNVA}, the ability to specify the requisite exposure model with understanding of the physical process dictating the interference mechanism highlights an important distinction between studies of network interference on social networks and those governed by complex exposure patterns.  In a social network context, the network structure is typically part of the data collection; network ties are recorded based on an explicit criterion for connection between units.  For example, two people are connected in the network if they report being friends. As a consequence, difficulties in defining and measuring connections, which may have implications for downstream analysis, can be considered as inherent features of the data collection process.  In contrast, studies of network interference governed by complex exposure patterns that depend on other physical processes specify a (physical or statistical) model for the network connections. Thus, any deficiency in the characterization of network connections is not part of data collection, but rather the specification for the mechanism generating interference.  This highlights the importance of incorporating, when available, extant knowledge of the mechanistic dynamics generating complex exposure dependencies. The threat of downstream analysis bias should be judged against the relative understanding of any supposed process dynamics.

As a consequence of Assumption \ref{ass:SUTNVA}, each outcome unit can be regarded as receiving a ``treatment" that is dictated jointly by two components, $Z_i$ and $G_i$.  The assignment to $Z_i$ is governed by the process that dictates whether the interventional units that are key associated to any outcome unit adopt treatment.  The assignment to $G_i$ is governed by the combination of the process that dictates whether {\it any} interventional unit adopts treatment and the structure of the interference network specified in $T$.  This leads to formalization of an assignment mechanism governing the joint treatment, denoted with
\begin{equation}
P(\vZ, \vG|\vX^{out}, \vX^{int}, \{\vY(z, g), z \in \{0,1\}, g \in \mathcal{G}\}), \label{eqn:assignment}
\end{equation}
where $g \in \mathcal{G}$ is, in a slight abuse of notation, taken to denote the values of $g$ that are contained in $\mathcal{G}_i$ for all $i$. \cite{forastiere_identification_2021} formulated a similar assignment mechanism in the case of one level of observational unit, but in a setting where $\vZ$ and $\vG$ were deterministically linked for a fixed $T$.  In contrast, the assignment mechanism in (\ref{eqn:assignment}) permits independent variation in the two components of treatment, even for a fixed $T$. This results from the fact that the vector $\vZ$ encodes the treatment statuses of only the interventional units that are key-associated to at least one outcome unit (i.e., the elements of $\vS$ corresponding to $\{ \jstar ; i=1,2,\ldots, n \}$), whereas $\vG$ derives from the entire vector $\vS$.  Thus, insofar as there are elements of $\vS$ contained in the calculation of $\vG$ but not represented in $\vZ$, it is possible, for a fixed $T$, for two different vectors of allocations to the interventional units, $\vS, \vS'$, to yield the same value of $\vZ$, but different values of $\vG$.   As a consequence, it is possible (and indeed relevant in the power plant analysis) to conceive of interventions that would vary the value of $\vG$ without changing the value of $\vZ$. This decoupling of $\vZ,\vG$ in the assignment mechanism has important implications for the interpretation of the causal estimands that will be presented in Section \ref{sec:estimands}.  

\subsection{Key-Associated Bipartite Causal Estimands}\label{sec:estimands} 
We define two causal estimands of interest that are anchored to the above definition of $Z_i$ and $G_i$, both motivated by common notions of ``direct'' and ``indirect'' effects pertaining to the effect of treating an individual unit and the effect of treating ``other'' units.  In the general bipartite setting, the lack of immediate delineation of which treatment applies ``directly'' to an outcome unit complicates the definition and meaning of such effects, but with the simplifications described in Section \ref{sec:potential_outcomes}, ``direct'' will be used in reference to the key associated unit, and ``indirect'' or ``upwind'' used in reference to all other units.  


In order to specify these causal estimands, we begin by reminding that the following quantities, all viewed as random variables, are associated with each outcome unit $i$,
$$ Y_i(z, g), Z_i \in \{0,1\}, G_i \in \mathcal{G}, X_i^{out}, X_i^{int}.$$
We are interested in the expected values of the potential outcomes under particular values of $Z_i=z$ and $G_i=z$ and conditional on specific values of the covariates, i.e.,  $X_{i}^{out}=x^{out}$ and $X_{i}^{int}=x^{int}$. Denote this expected value as:
\begin{equation}
    \mu_x(z, g)\equiv E[ Y_i(z, g)\mid X_{i}^{out}=x^{out}, X_{i}^{int}=x^{int}]. \label{def:CATE}
\end{equation}
The causal estimands of interest will be based on $\mu_x(z,g)$, marginalized over the empirical distribution of covariates in the finite sample defined by all ZIP codes and coal-fired power plants in the United States during 2005. Formally:
\begin{equation}
\label{eq:mu}
\mu(z,g)=\E_{X^{int}, X^{out}}[\E_{Y(\cdot)| X^{int}, X^{out}}[Y_i(z,g)| X_i^{int}, X_i^{out}]]
\end{equation}
where the expectation $\E_{X^{int}, X^{out}}(\cdot)$ is over the empirical distribution of covariates in our population of interest.
In summary, $\mu(z,g)$ represents the expected value  of the potential outcome under key-associated treatment $z$ and upwind treatment $g$ for a representative unit of the finite sample of interest.



This model-based approach implicitly assumes that the value $Y_i(Z_i=z, G_i=g)$ is well defined for all outcome units.  We continue development under this assumption for ease of exposition and because the interference process considered in the power plant example supports this assumption, but note that \cite{forastiere_identification_2021} explicitly considers the possibility that the structure of the network would render certain values of $G_i$ impossible for some $i$.

Using the above notation for the average potential outcome, we first define an estimand akin to an average ``direct effect'' of treating the key associated unit while holding fixed the treatments of other units:
\begin{equation}
\tau(g) = \mu(1,g) - \mu(0,g)
\label{eq:tau}
\end{equation}
which might correspond, for example, to the average effect on IHD hospitalizations of installing (vs. not) a scrubber on the most influential power plant, while holding fixed the scrubber statuses of all upwind plants, or at least their HyADS-weighted scrubber rate $G_i=g$ .  An average direct effect over the distribution of $G$ can be defined with $\tau = \sum_{g \in \mathcal{G}} \tau(g)P(G_i=g)$, where $\mathcal{G} = \cup_i \mathcal{G}_i$.

Another estimand, akin to an ``indirect'' or ``spillover'' effect, can be defined as:
\begin{equation}
\delta(g;z) = \mu(z,g) - \mu(z,g^{min})
\label{eq:delta}
\end{equation}
to denote the average effect of changing the treatment statuses of interventionial units in the interference mapping to toggle the ``upwind'' treatment from $g$ to $g^{min}$, where $g^{min}$ could be defined as any relevant value of $G$, for example, the minimum value observed in the sample.  In keeping with the interpretation of the power plant example, we will refer to this effect as the ``upwind'' effect, interpretable as the average effect on IHD hospitalizations of having upwind scrubber rate $g$ relative to the smallest realistic upwind scrubber exposure, while holding fixed the scrubber status of the most influential plant at $z$. An average upwind effect over the distribution of $G$ can be defined as $\Delta(z) = \sum_{g \in \mathcal{G}} \delta(g; z)P(G_i = g)$.  

The estimands in (\ref{eq:tau}) and (\ref{eq:delta}) are based on setting both the treatment of the key associated interventional unit and the treatments of interventional units in the interference mapping. Average direct and upwind effects are then calculated according to the (empirical) distribution of $P(G_i=g)$.  These average estimands are similar to the ones introduced in \cite{forastiere_identification_2021}, and isolate the effect of a specific intervention on a key-associated interventional unit from the effect of
changing the distribution of the treatment in the population of interventional units (see  \cite{forastiere_identification_2021}, Section 2.4, for a discussion). This stands in contrast to other work that defines average effects over hypothetical interventions on the whole sample, (e.g. general stochastic interventions
in \cite{van_der_laan_causal_2014} and its extensions or Bernoulli trials in \cite{liu_inverse_2016}) or the spatial stochastic interventions in \cite{papadogeorgou_causal_2022}.

\subsection{Ignorable Treatment Assignment}
With the bivariate treatment formulated above and the corresponding joint treatment assignment mechanism, identification of causal effects relies on an assumed version of ignorable treatment assignment.  With an abuse of notation to let $j \in t_i^\top$ denote the set $\{j; t_{ij} >0\}$, we state:
\begin{assumption}[Ignorability of Joint Treatment  and Confounding in the Bipartite Setting]
\label{ass:ignorability}
\[
Y_i(z,g) \ind Z_i,G_i \mid  \{\vX^{int}_{j}\}_{j \in t_i^\top}, \vX^{out}_i \qquad \forall z\in \{0,1\}, g
\in \mathcal{G}_i, \forall i.
\]
This assumption states that the treatment status of an outcome unit's key-associated interventional unit and that outcome unit's upwind treatment are independent of the potential outcomes, conditional on a the covariates for the interventional units in the $i^{th}$ unit's interference set ($\{\vX^{int}_{j}\}_{j \in t_i^\top}$) and the covariates of the outcome unit ($\vX^{out}_i$).
\end{assumption}

Assumption 2 does not pertain to the entire assignment mechanism in (\ref{eqn:assignment}), but rather to the relationships among treatments and potential outcomes at the individual unit level.  Thus, it may hold irrespective of any dependence between the treatment assignments to interventional units or between the potential outcomes of different outcome units.  The set of confounders $\{\vX^{int}_{j}\}_{j \in t_i^\top}$ and $\vX^{out}_i$ that should be conditioned on to satisfy Assumption \ref{ass:ignorability} include all those covariates related to either $Z_i$ or $G_i$ and the outcome $Y_i(z,g)$. This requires careful consideration in the bipartite case, as it introduces novel notions of what might be called  ``neighborhood confounding'' corresponding to the different notions of how units may interfere described in Section \ref{sec:treat_functions}.
 
The basic notion for interfering outcome, interventional units $(i,j)$ introduces two different types of confounding indicated in Assumption \ref{ass:ignorability}.  First, consider covariates in $\{\vX^{int}_{j}\}_{j \in t_i^\top}$, denoting the covariate vectors for interventional units that interfere with outcome unit $i$. Confounding arises when these interventional features relate to outcomes at unit $i$ and to the key-associated and/or neighborhood treatments, the latter arising if elements in $\{\vX^{int}_{j}\}_{j \in t_i^\top}$ dictate the adoption of treatments encoded by $\vS$.  We refer to this as {\it upwind (or upstream) confounding.} An example of an upwind confounder in the power plant case would be where $\{\vX^{int}_{j}\}_{j \in t_i^\top}$ denotes the heat input of all power plants upwind to location $i$, which could relate to $(Z_i, G_i)$ if larger power plants with higher heat input are more likely to install scrubbers and if larger power plants tend to be located near population centers exhibiting other features (not captured in $\vX^{out}_i$) that dictate hospitalization rates.  As a practical matter, summary functions of the features in $\{\vX^{int}_{j}\}_{j \in t_i^\top}$ may be required to avoid adjusting for a high-dimensional set of features at many  interventional units in unit $i$'s interference set.


Next, consider covariates in $\vX^{out}_i$, denoting features of outcome unit $i$. Confounding arises when these features relate to the outcome at unit $i$ and the treatment assignments in $\vS$ that dictate the value of $(Z_i, G_i)$. Dependence between $\vX^{out}_i$ and $(Z_i, G_i)$ could arise if treatments at interventional units are impacted by features of interfering outcome units, a phenomenon we refer to as {\it downwind (or downstream) confounding}.  For example, in the power plant setting, decisions to install scrubbers may be based in part on knowledge of downwind population centers or particular areas cited for regulatory noncompliance, which may also exhibit certain patterns of hospitalization. For practical purposes, it may be required to encode this information at each interventional unit with some summary feature of the outcome units in its interference set, $h(\{\vX^{out}_{i}\}_{i \in t_j^\top})$.  For example, if treatment decisions at power plants are informed by population density of downwind ZIP codes,  $h_j(\cdot)$ could  denote the average population density of all ZIP codes downwind from plant $j$ and included within $\vX^{int}_j$.

The possibility of overlapping interference sets introduces a third type of confounding.  Since, in general, interventional units interfere with multiple outcome units,  downwind confounding would imply dependence between ($Z_i, G_i$) and the covariates of all other outcome units that interfere with any $j \in t_i^\top$. Consider ZIP codes $(i, i')$ that have overlapping interference sets that share power plant $j$. In the presence of downwind confounding, treatment adoption at unit $j$ may depend on covariates at both $i$ and $i'$. Thus, $(G_i, Z_i)$ will depend on the confounder values of unit $i'$ (and {\it vice versa}). Confounding by characteristics of interventional units with overlapping interference sets could be defined analogously.

In the power plant case, confounding due to overlapping interference sets relates to the potential for confounding due to the notion of {\it homophily}, that is, the tendency of nodes with similar features to share edges. The closest analogue to homophily in the bipartite setting is outcome units with similar features tending to share connections with similar sets of interventional units. This could arise because units with overlapping interference sets are likely to be spatially close, and thus share similar features, inducing a joint association between $(G_i, Y_i(z,g))$ and potential outcomes at nearby outcome units. For example, a collection of ZIP codes with high population density may impact a nearby power plant's decision to install a scrubber (i.e., downwind confounding). If these ZIP codes are spatially close (or clustered), then their similarity in population density may also coincide with similarity in potential hospitalization rates. A related phenomenon in the explicitly spatial power plant setting is the possibility spatial correlation of potential outcomes.  The spatial nature of $G_i$ could imply that its value for an index ZIP code is related to the potential outcomes of nearby ZIP codes which, when potential outcomes are spatially correlated, could yield confounding due to outcomes at nearby ZIP codes being jointly associated with $Y_i(z,g)$ and $G_i$.  The threat of confounding due to homophily or spatial correlation can be mitigated by recognition of whether the underlying reasons for shared edges in the interference network are fully specified (e.g., as in a physical process with the HyADS model), and thus not a consequence of unobserved similarities among units that often threaten the validity of studies of social networks. In the case of our power plant investigation, it is relevant that HyADS is not equivalent to spatial proximity; hence similar values in the HyADS matrix do not imply ZIP codes are geographically close and share similar demographic features.  The notation of Assumption \ref{ass:ignorability} implies that any confounders from outcome units with overlapping interference sets are encoded in $\{\vX^{int}_{j}\}_{j \in t_i^\top}$ (possibly via summary functions) and that the threat of residual confounding due to spatial correlation is mitigated by inclusion of $\vX^{out}_i$ that are themselves spatially correlated in accordance with the spatial patterning of the outcome.   We describe specific assumptions about the threat of these types of confounding in the Power Plant analysis in Section \ref{sec:analysis}.

\cite{forastiere_identification_2021} show that, in combination with the assumption of consistency, Assumptions \ref{ass:SUTNVA} and \ref{ass:ignorability} are sufficient to identify the causal effects from Section \ref{sec:estimands}.  In addition, we adopt a version of the positivity assumption that the joint treatment probability in (\ref{eqn:assignment}) is strictly between 0 and 1, that is, there is no combination of $\vX_i^{out}, \{\vX^{int}_{j}\}_{j \in t_i^\top}$ that deterministically dictates the individual or upwind treatment, which is of practical importance when confronting issues of in-sample overlap between the covariate distributions of units with different levels of the observed key-associated and upwind treatment. Ignorability under spatial interference is also discussed in some detail in \citet{zirkle_addressing_2021}, but focusing only on the individual treatment in the non-bipartite setting.

\section{Estimating Bipartite Causal Effects with Joint Propensity Scores} \label{sec:gps}
For estimating the causal effects defined in Section \ref{sec:estimands}, we adopt an estimation approach similar to that in \cite{forastiere_identification_2021}, which estimates a joint propensity score that, under Assumption \ref{ass:ignorability}, has similar properties and can be used in a manner similar to how propensity scores have been previously adopted to adjust for confounding in observational studies \citep{rosenbaum_central_1983, imbens_role_2000, hirano_propensity_2004, stuart_matching_2010}. \cite{giffin_generalized_2020} and \cite{zirkle_addressing_2021} also exploit the use of generalized propensity scores in settings of spatial interference; the former includes a bivariate version in a smoothed Bayesian model and the latter focuses only on the individual propensity score and using it for design purposes (matching, trimming).

Under Assumption \ref{ass:ignorability}, identification of causal effects follows from a joint propensity score that represents, for each outcome unit, the marginal probability of receiving treatment $(Z_i, G_i)$, construed as an average probability over the outcome units having the same covariate values (\cite{imbens_causal_2015}, pg. 34-35).    As in \cite{forastiere_estimating_2022} and \cite{forastiere_identification_2021}, this approach relies on the common premise that the propensity score serves to summarize covariate information contained in the sample such that conditioning on the summary renders the mechanism assigning treatments to units ignorable \citep{ho_matching_2007, imbens_causal_2015, rubin_use_1985}. Thus, we do not pursue a model for the joint distribution of $(Z_i, G_i |  \vX_i^{out}, \{\vX^{int}_{j}\}_{j \in t_i^\top})$ across all $i$ that is fully consistent with the entire assignment mechanism in (\ref{eqn:assignment}): the goal of the joint propensity score strategy is to balance the relevant features in  $(\vX_i^{out}, \{\vX^{int}_{j}\}_{j \in t_i^\top})$ in order to adjust for confounding by features that differ across outcome units with different values of $(Z_i, G_i)$. 

\subsection{Specification of the Joint Propensity Score}
The {\it joint propensity score} governing the assignment of the key-associated and upwind treatments can be denoted as
\begin{equation}
\begin{aligned}
\psi(z, g; x^{int}, x^{out})&=P(Z_i=z, G_i=g|  \{\vX^{int}_{j}\}_{j \in t_i^\top}=x^{int}, \vX_i^{out}=x^{out})\label{eq:jointps}\\
\end{aligned}
\end{equation}
\cite{forastiere_identification_2021} established several properties of the joint propensity score in (\ref{eq:jointps}) which carry over to the bipartite setting under the above definitions of the key-associated and upwind treatment and Assumptions \ref{ass:SUTNVA} and \ref{ass:ignorability}. First, \cite{forastiere_identification_2021} established the balancing property of (\ref{eq:jointps}), implying that, among outcome units with the same value of $\psi(z,g; x^{int}, x^{out})$, the distribution of $(\{\vX^{int}_{j}\}_{j \in t_i^\top}, \vX^{out})$ is the same between units with $Z_i=z, G_i=g$ and units with other values of $(Z_i, G_i)$. This property, combined with Assumption \ref{ass:ignorability}, implies that $Y_i(z,g) \ind Z_i,G_i \mid  \psi(z,g; x^{int}, x^{out})$ for $ z\in \{0,1\}, g
\in \mathcal{G}_i, \forall i$, i.e., that is, that potential outcomes are independent of the key-associated and upwind treatments, conditional on the joint propensity score. Therefore, it is sufficient to adjust the joint
propensity score to account for confounding bias in the estimation of both direct and spillover
effects.

Consequently, \cite{forastiere_identification_2021} show that average potential outcomes in (\ref{eq:mu}) are identified as a function of the observed data as $E[E[Y_i|Z_i=z, G_i=g, \psi(z, g; x^{int}, x^{out})]|Z_i=z, G_i=g]$, 
where the outer expectation is  over the distribution of the joint propensity score and the inner expectation is over the distribution of observed outcomes.  Note that this result applies with any balancing score.

As a practical matter, the multi-valued nature of the joint treatment (owing to the scale of the upwind treatment, $G$) makes direct adjustment for $\psi(z, g; x^{int}, x^{out})$ difficult.  However, the binary nature of the key-associated treatment motivates the following factorization of the joint propensity score:
\begin{align}
\psi(z, g; x^{int}, x^{out}) =&P(Z_i=z, G_i=g| \{\vX^{int}_{j}\}_{j \in t_i^\top}=x^{int}, \vX_i=x^{out}) = \nonumber \\
&P(G_i=g| Z_i=z, \{\vX^{int,g}_{j}\}_{j \in t_i^\top}=x^{int,g}, \vX^{out, g}_{i} = x^{out,g}) \times \label{eq:facg}\\
&P(Z_i=z|\{\vX^{int,z}_{j}\}_{j \in t_i^\top}=x^{int,z}, \vX^{out,z}_i = x^{out,z}) \label{eq:facz}
\end{align}
where we denote the part of the factorization in (\ref{eq:facg}) with $\lambda(g; z, x^{int,g}, x^{out,g})$ to represent the probability of having upwind treatment level $g$ conditional on $z$ and covariates, and we denote the part of the factorization in (\ref{eq:facz}) with $\phi(z;x^{int,z}, x^{out,z})$ to denote the probability of having the key-associated treatment at level $z$, conditional on covariates. 
$\lambda(g; z, x^{int,g}, x^{out,g})$ will be referred to as \textit{upwind propensity score}, while $\phi(z;x^{int,z}, x^{out,z})$ will be referred to as \textit{key-associated propensity score}.
Note the expanded notation to reflect the possibility of refining the model specification with the assumption that these two components of the joint propensity score model might not include the same covariates; $\vX^{out, g}$ represents the outcome-unit covariates relevant to the assignment of $G$ and $\vX^{out,z}$ represent the outcome-unit covariates relevant to the assignment of $Z$, with analogous definitions for $\vX^{int, g}$ and $\vX^{int, z}$.  The binary nature of the key-associated treatment, $Z$ (relating to $\phi(z;x^{int,z}, x^{out,z})$), supports for the key-associated propensity score the use of techniques common to the literature on propensity score adjustment for binary treatments, while because of the continuous domain of the upwind treatment, $G$ (relating to $\lambda(g; z, x^{int,g}, x^{out,g})$), the upwind propensity score can be viewed in the way generalized propensity scores have been proposed in the context of continuous treatments \citep{hirano_propensity_2004}. 
For this reason, throughout we will use the terms upwind propensity score and generalized propensity score (GPS) interchangeably. 
This result forms the basis for an unbiased estimator over repeated sampling from the potential outcome model (Equation \eqref{def:CATE}) conditional on units' realized covariates and repeated randomizations from the assignment mechanism.   

Because \citet{forastiere_identification_2021} proved that  the (marginal) joint propensity score retains the properties of a balancing score and can adjust for confounding, our proposed estimator, outlined in next Section, uses an estimated (marginal) GPS, a model for Y specified conditional on the estimated GPS, and stratification on the individual propensity score. The particulars of these modeling specifications may lead to some finite sample bias due to residual imbalance after stratification and model misspecification.
We mitigate the bias by evaluating the balancing properties of the estimated GPS: if balancing is achieved the bias should be negligible.
Any bias due to  residual imbalance, stratification and model misspecification in our scenarios will be evaluated via simulations (Section 5). A bootstrap procedure is also proposed (Section 4.3) to recover sampling and assignment uncertainty, and its performance is also evaluated via simulations.
\subsection{Estimating Procedure: Subclassification and Generalized Propensity Score Adjustment}\label{sec:est.strategy}
We outline one approach for confounding adjustment with the joint propensity score that unfolds in two steps. 
First, estimates of the key-associated propensity score, $\phi(z;x^{int,z}, x^{out,z})$ are obtained from a model of the form $P(Z_i=z|\{\vX^{int,z}_{j}\}_{j \in t_i^\top}, \vX_i^{out,z}) = f^Z(z,\{\vX^{int,z}_{j}\}_{j \in t_i^\top}, \vX_i^{out,z};\gamma)$, with predicted values from this model, denoted with $\hat{\phi}_i$ for each of $i=1,2,\ldots,n$. Then, estimates of the upwind propensity score are obtained with a parametric model, $\lambda(g; z, x^{int,g}, x^{out,g})$, specified as $P(G_i =g|Z_i=z, \{\vX^{int,g}_{j}\}_{j \in t_i^\top}, \vX^{out,g}_i) = f^G(g,z,\{\vX^{int,g}_{j}\}_{j \in t_i^\top}, \vX^{out,g}_i; \delta)$. Density estimates from this model, denoted with $\hat{\lambda}_i$, are estimates of the upwind propensity score. 
Specification of the key-associated and upwind propensity scores can by judged by assessing the balancing property of the estimated $\hat{\phi}_i$ and $\hat{\lambda}_i$.

After estimating $\hat{\phi}_i$ and $\hat{\lambda}_i$, each of the $n$ outcome units is then assigned to one of $K$ strata, denoted $K_1, K_2, \ldots, K_K$, based, for example, on quantiles of the $\hat{\phi}_i$.  The covariates $\{\vX^{int,z}_{j}\}_{j \in t_i^\top}, \vX^{out,z}$ should be balanced between units with $Z=0$ and those with $Z=1$ within each of the $K$ strata, which can be checked empirically. The observed data within each of the $K$ strata are then used to estimate a model for the potential outcomes $Y_i(z,g) | \hat{\lambda}_i \sim f^y(z,g,\hat{\lambda};\theta_k)$.  Predicted values from this model, denoted with $\hat{Y}_i(z,g)$, represent estimated potential outcomes for every level of $(Z_i=z, G_i=g)$ across a pre-defined grid of values. The estimated within-stratum dose-response function $\mu_k(z,g)$ is then obtained by averaging these predicted potential outcomes for each value of $(z,g)$ as:
\[\widehat{\mu}_k(z,g)=\frac{\sum_{i\in n_k} \widehat{Y}_i(z,g)}{n^k}.\]
These within-stratum dose-response functions are then averaged over the $K$ strata to obtain an overall estimate with
$\widehat{\mu}(z,g)=\sum_{k=1}^K \widehat{\mu}_k(z,g) \pi^k$, where $\pi^k$ denotes the proportion of observations observed to lie in stratum $K_k$.  Estimates of $\widehat{\mu}(z,g)$ are then used to obtain estimates of the causal estimands defined in Section \ref{sec:estimands}.

\subsection{Interventional-Unit Bootstrap Approximation}
For inference, we adopt a novel
bootstrap approach designed specifically to
recover the complex correlation structure resulting from the mechanism that assigns treatments to interventional units ($\vS$), which, for fixed T, generates assignments $\vZ$ and $\vG$ to outcome units, along with the variability owing to the sampling of potential outcomes. Our resampling procedure is characterized by the following key steps: 1) we resample interventional units with replacement; 2) 
we preserve the entire network structure $T$ across bootstrap samples, including the key-associated unit ($\jstar$) for each outcome unit; 3) we retain all outcome units that are key-associated to the sampled interventional units in each bootstrap sample; 4) for each outcome unit $i$ in the bootstrap sample, in addition to their observed outcome $Y_i$ and their outcome-unit covariates 
$X_i^{out}$, we preserve their key-associated treatment $Z_i$, as well as their upwind treatment $G_i$ and their interventional-unit covariates $\{\vX_i^{int}\}_{j \in t_i^{\top}}$ that are computed based on the entire sample of interventional units, regardless of the ones that are included in the bootstrap sample 
\footnote{The latter approach is similar in spirit to the ``egocentric'' approach proposed by \citet{forastiere_identification_2021}.}.
Because we resample interventional units as opposed to outcome units, we name this type of bootstrap approach \textit{interventional-unit bootstrap}. It is worth noting that, while this type of resampling of interventional units in step 1) is designed to reflect the assignment of $\vS$ to interventional units, step 3) is designed to preserve the correlation structure in the key-associated treatments $\vZ$, whereas step 4) would partly recover the correlation structure in $\vG$, thanks to the complex spatial structure induced by $T$ and the correlation between outcome units with the same key-associated power plant.


An obvious alternative could be resampling the outcome units and retaining their key-associated and upwind treatment as well as their covariates constant across bootstrap samples, as proposed and evaluated in \cite{forastiere_identification_2021}). 
This resampling procedure assumes that the observed data, including the key-associated and upwind treatments, is independent across outcome units. 
However, the present bipartite setting significantly deviates from this independence assumption, as network structure $T$ is dense and many units have overlapping interference sets. In addition, in the present power plant setting we have a number of interventional units much lower than the number of zip codes, i.e.,  $J=472$ and $N=\numzips$, which leads to a highly correlated treatment structure. Intuitively, an outcome-unit bootstrap can be seen as `oversampling power plants' in each bootstrap sample, as $Z_i$ and $G_i$ are retained for each independently sampled outcome unit.
The simulation study in Section \ref{sec:sim} presents this outcome-unit bootstrap for comparison with our proposed interventional-unit bootstrap.

To the extent that the interventional-unit bootstrap that fixes node characteristics and retains only the key-associated outcome units for every resampled interventional unit is an approximation for how the data (e.g., in the power plant investigation) are presumed to be generated, the validity of the bootstrapped variance estimates is not guaranteed, but when evaluated in realistic simulation scenarios in Section \ref{sec:sim} that mimic the structure of the problem and the actual observed data, is shown to be conservative for the variance that would arise with respect to repeated sampling of potential outcomes from the corresponding models and repeated assignment of treatments to the interventional units..




\section{Simulation Study}\label{sec:sim}
We offer a simple simulation study to evaluate the operating characteristics of the proposed joint propensity score estimation approach in data-generating scenarios meant to mimic the realities of the power plant investigation. The data generation fixes the outcome units to be the $N=\numzips$ ZIP codes retained for the power plant analysis in Section \ref{sec:analysis}, the interventional units to be the actual $J=472$ power plants, and $T$ to be that specified with the actual HyADS matrix described in Section \ref{sec:hyads}. Random generation of treatment assignments to power plants from an assignment mechanism and outcomes measured at outcome units from a specified model generate variability in the observed outcomes in a finite population of interest.  Thus, major goals of this simulation study are to illustrate whether reasonable model specifications for the joint propensity score can adjust for confounding and provide (approximately) unbiased estimates of the direct and upwind effects and whether the interventional-unit bootstrap provides a reasonable approximation of uncertainty, albeit with an estimation procedure that does not directly correspond to the presumed underlying data-generating mechanism. 

\subsection{Data Generating Process}
  We consider  $\vX^{int}_{j}$ to include two interventional-unit characteristics: percent operating capacity (\textit{Capa}) and (log-transfored) heat input (\textit{Heat}). Also included in $\vX^{int}_{j}$ is a downwind (or downstream) confounder, calculated as the the HyADS weighted average population in ZIP codes downwind from each power plant:  \(down.pop_j = \sum_{i} t_{ij}\times Pop_i \) (rescaled  to have mean 0 and variance 1).  $\vX^{out}_{i}$ is the (log transformed) total population for ZIP code $i$ (\textit{Pop}). Treatment assignments for each of the $J=472$ power plants corresponding to the installation of scrubbers are simulated as independent Bernoulli random variables with probability of treatment specified as follows to depend on the two power plant characteristics and the downstream population confounder:
\begin{equation}\label{eq:upw_pop_sj}
\mathrm{logit}(P(S_j = 1)) = 1.2 \times Capa_{j} + 0.15 \times Heat_{j} + 0.15 \times down.pop_j - 3).
\end{equation}

The key-associated and upwind treatment are then calculated based on the HyADS matrix ($T$) to set  $Z_i=S_{\jstar}$ and $G_i = \sum_{j \ne \jstar}t_{ij}S_j$ where, as in Sections \ref{sec:treat_functions}, and \ref{sec:analysis}, $\jstar$ denotes the power plant key-associated to the i$^{th}$ ZIP code based on the value of $t_i$ with the highest value ($j; t_{ij} = \max_j \{t_i\}$).

We simulate outcomes \(Y_i\) from a normal distribution centered at \(\mu_i\) with standard deviation 1.   The mean $\mu_i$ is specified to depend directly on the joint treatment, the outcome-unit covariate, and the covariates of the key-associated interventional unit.  Specifically:
\begin{equation}
\mu_i = 5 \times Capa_{j^*(i)} + Heat_{j^*(i)} + Pop_i - Z_i - G_i
\end{equation}
This data generation implies a true value of the direct effect $\tau = -1$.  To evaluate the operating characteristics of the estimators, we calculate the true value of the upwind effect using simulations. For each replicate data set, the true value of the upwind effect is calculated using the simulated potential outcomes for that data set, with the average true upwind effect across all replicate data sets taken as the true population value.  
 
We generate 200 replicate data sets by generating new vectors $\vS$ (corresponding to new $(\vZ, \vG)$) and new outcomes $Y_i$.  For each replicate data set we implement the proposed generalized propensity score-based estimator.  The key-associated propensity scores, $\phi(z;x^{int,z}, x^{out,z})$, are estimated with $f^Z(z,\{\vX^{int,z}_{j}\}_{j \in t_i^\top}, \vX_i^{out,z};\gamma)$ specified as a logistic regression with  $\mathrm{logit}(P(Z_i = 1)) = Capa_{j^*(i)} + Heat_{j^*(i)} + Pop_i$. The upwind propensity scores, $\lambda(g; z, x^{int,g}, x^{out,g})$, are estimated with $f^G(g,z,\{\vX^{int,g}_{j}\}_{j \in t_i^\top}, \vX^{out,g}_i; \delta)$ specified as a linear regression on $Z_i + Capa_{j^*(i)} + Heat_{j^*(i)} + Pop_i$. 

\subsection{Estimation}
For estimating causal effects, units are subclassified into quintiles based on estimates of the key-associated propensity score.  Within each subclass, potential outcomes are modeled with a linear regression model.  $Y_i$ is regressed on $Z_i, G_i, \hat{\lambda}_i$, and $G_i \times \hat{\lambda}_i$, where the interaction term is included to provide additional flexibility in light of the unknown dependence between $Y_i$ and the joint propensity score. An unadjusted analysis that simply regresses $Y_i$ on $(Z_i, G_i)$, without stratification, is included for comparison and to illustrate the degree of confounding in the direct and upwind effects. For the propensity score  based methods, standard errors for estimated direct and upwind effects are estimated with the interventional-unit bootstrap (based on 100 bootstrap samples) and, for comparison, the outcome-unit bootstrap, both described in Section \ref{sec:est.strategy}.

\subsection{Relationship Between Presumed Data Generation and the Joint Propensity Score Model}
There are several points worth noting about the relationship between the above data generation and the models used for estimating causal effects that are specific to the bipartite setting.  First is the role of the donwstream confounder (\(down.pop_j\)) when generating the data.  Scrubber installations depend on \(down.pop_j\) to reflect that a power plant may adopt treatment based on knowledge of the downwind population.  However, note that \(down.pop_j\) does not directly appear in the generation of outcomes (or in the model for the joint propensity score): $Y_i$ is generated based on \(Pop_i\) to more realistically reflect that an outcome at outcome unit $i$ might depend only on the population at that location, even though  \(Pop_i\) bears only an indirect relationship with the downstream confounder \(down.pop_j\) used to generate treatments.  

Second, note the discrepancy between the data generation for the joint treatment and the specification of the joint propensity score model. The treatment generation results from random simulation of $\vS$ based on covariates, whereas the joint propensity score model used for estimation directly models the dependence between $(Z_i, G_i)$ and covariates. This is a key feature of the bipartite data generation; values of $(Z_i, G_i)$ are implied by the combination of treatment assignments to interventional units ($\vS$) and the specified adjacency matrix ($T$), where outcome units with the same key-associated interventional unit are constrained to have the same key-associated treatment.  Nonetheless, models for the joint propensity score are specified directly on $(Z_i, G_i)$ to estimate the marginal propensity score for each outcome unit. 


\subsection{Simulation Results}
Table \ref{table:sim} summarizes the performance of the proposed estimation approach in these data generations.  The average bias in unadjusted estimates of direct and indirect causal effects indicate the threat of confounding in these data generations, while the proposed joint propensity score estimates appear close to unbiased.  The conservative nature of the interventional-unit bootstrap is clearly evident, particularly for the upwind effects, with average bootstrap standard errors well exceeding the empirical standard error of the point estimates, in turn leading to coverage above the nominal level.  Thus, this simulation study indicates that, even with the noted discrepancies between data generation and model specification, the proposed joint propensity method can recover unbiased estimates of causal effects, with the interventional-unit bootstrap conservative for the variance that arises with respect to sampling outcomes from the model  and  assignment of treatments to interventional units. For comparison, Table \ref{table:sim} shows the that the outcome-unit bootstrap that resamples ZIP codes yields standard error estimates that are far too small, leading to very poor coverage.

\begin{table}[ht]
\centering
\begin{tabular}{r|c|cccccc}
  \hline \hline
    & Unadjusted &  \multicolumn{6}{c}{Joint Propensity} \\
& & & & \multicolumn{2}{c}{Interventional Unit}& \multicolumn{2}{c}{Outcome Unit}\\
 & & & & \multicolumn{2}{c}{Bootstrap} & \multicolumn{2}{c}{Bootstrap} \\
 & Bias & Bias & Empirical SE & Avg. SE & 95\% Coverage & Avg. SE & 95\% Coverage \\ 
  \hline
$\tau$ & 0.45 & 0.05 & 0.10 & 0.15 & 0.99 & 0.028 & 0.320\\ 
  $\Delta(0)$ & -0.60 & -0.01 & 0.61 & 1.16 & 1.00 & 0.168 & 0.385\\ 
  $\Delta(1)$ & -0.60 & -0.01 & 0.61 & 1.17 & 1.00 & 0.169 & 0.385\\ 
   \hline
\end{tabular}
\caption{Simulation study: Bias and Standard Error (SE) based on 200 monte carlo replicates analyzed with an unadjusted model and the proposed joint propensity score approach with 100 bootstrap samples.}\label{table:sim}
\end{table}

\section{Evaluating the Effectiveness of scrubbers for reducing Medicare IHD Hospitalization}\label{sec:analysis}
We deploy the HyADS dispersion model and statistical methods described in the previous sections to evaluate the extent to which presence of scrubbers on coal-fired power plants in 2005 caused improvements in IHD hospitalizations among Medicare beneficiaries during that same year, accounting for the interference arising due to long-range pollution transport, as described in Section \ref{sec:background}.  

\begin{figure}
    \centering
    \includegraphics[width = \textwidth]{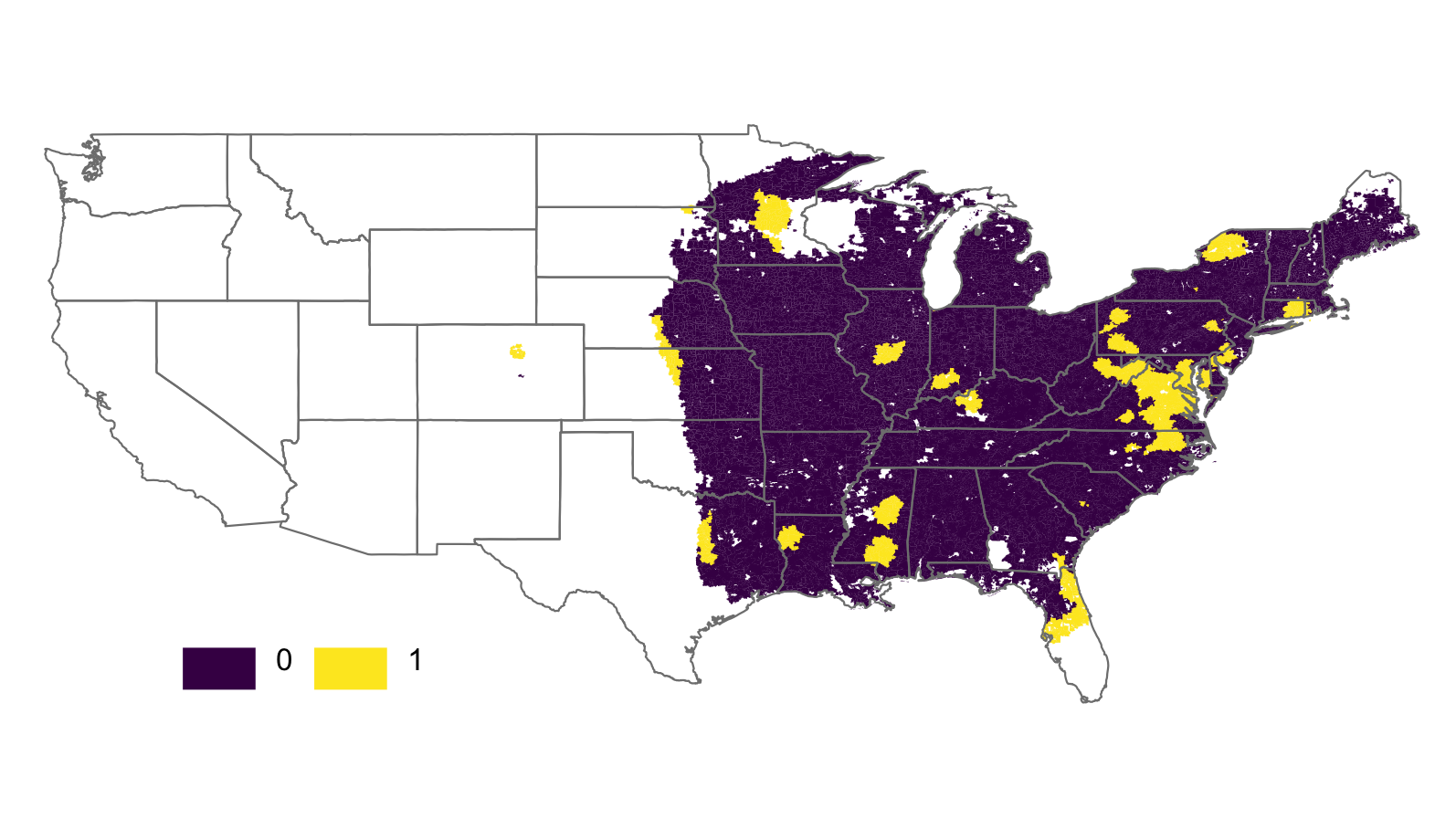}
    \caption{\numzips ZIP codes subject to meaningful coal power plant pollution in 2005, colored according to whether the key-associated plant has a scrubber ($Z_i=0$ or $1$). White areas are omitted from the analysis because of low power plant exposure or lack of propensity score overlap.}
    \label{fig:locations_map}
\end{figure}

Specifically, the interventional units are $J=\numplants$ electricity generating facilities (``power plants'') operating in 2005 that use coal as the primary fuel, of which \numscrubbers had scrubbers installed ($S_j=1$) during at least half of 2005. The HyADS approach of Section \ref{sec:background} was used to quantify the annual impact of air originating at each of the \numplants facilities on each ZIP code in the US.  For the analysis, $n=\numzips$ ZIP codes, lying mostly in the Eastern US (where most coal power plants are located) were retained on the basis of having an annual HyADS value in excess of the 25th percentile of the national distribution (that is, annual HyADS value greater than 2.066212) and meeting propensity score overlap criteria described later.  Thus, the outcome units are these $n=\numzips$ ZIP codes where pollution from coal-fired power plants comprises an important feature of the ambient air quality and where overlap was satisfied with respect to the key-associated propensity score. Figure \ref{fig:locations_map} presents a map of these ZIP codes, which contain data on \numbeneficiaries fee-for-service Medicare beneficiaries in 2005.

The same output from the HyADS simulations was used as the interference mapping, $T$.  Specifically, let $t_{ij}$ from Section \ref{sec:interference_map} be the value from the source-receptor matrix output from HyADS denoting how much air mass originating at power plant $j$ travels to ZIP code $i$.  The key-associated plant for each ZIP code $i=1,2,\ldots,\numzips$ was identified based on the plant exhibiting the highest HyADS influence on ZIP code $i$ during 2005, that is, $\jstar$ is the element of $t_i$ with the highest value ($j; t_{ij} = \max_j \{t_i\}$). In total, \numkeyassoc of the \numplants power plants were key-associated to at least one ZIP code, with \numkeyassocscrubbed of the these key-associated plants having scrubbers installed for at least half of 2005. This leads to a key-associated treatment, $Z_i = S_{\jstar} = 1$ for \numzipswithkeytreated ZIP codes whose most influential power plant had a scrubber for at least half of 2005.  Figure \ref{fig:locations_map} denotes which locations have $Z_i = 1$. 

\begin{figure}
    \centering
    \includegraphics[width = \textwidth]{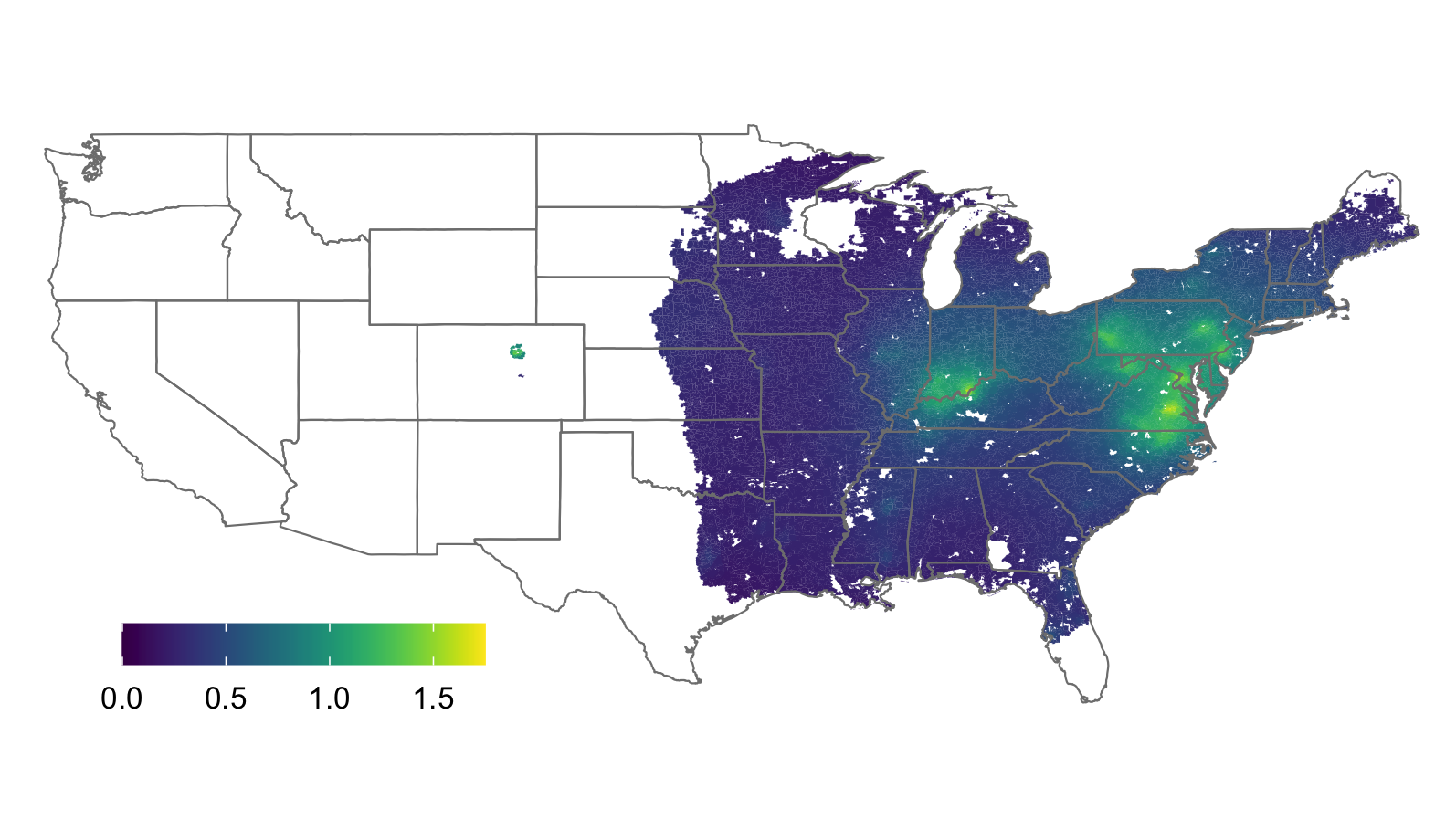}
    \caption{\numzips ZIP codes subject to meaningful coal power plant pollution in 2005, colored according to the HyADS-weighted upwind treatment rate, $G$.}
    \label{fig:locations_G}
\end{figure}

As stated in Section \ref{sec:treat_functions}, the upwind treatment, $G_i$ for $i=1,2,\ldots, \numzips$ is defined as a linear function of the treatment statuses of all power plants but power plant $\jstar$, weighted by the elements of the interference mapping: $G_i = \sum_{j \ne \jstar}t_{ij}S_j$.  Since $t_{ij}$, as output from HyADS, denotes the strength of influence of the $j^{th}$ interventional unit on the $i^{th}$ outcome unit, this function can be loosely interpreted as an ``upwind weighted'' rate of scrubbers among all but the key-associated power plant, so that a ZIP code can have high values of $G_i$ if it is heavily exposed to emissions from many power plants with scrubbers or from a few very influential power plants with scrubbers (or both). Figure \ref{fig:locations_G} maps the values of $G_i$ across the study area.

\subsection{Model specification for the joint propensity and potential outcomes}\label{sec:model_spec}
The covariates included in $\{\vX^{int,z}_{j}\}_{j \in t_i^\top}, \vX_i^{out,z}, \{\vX^{int,g}_{j}\}_{j \in t_i^\top}, \vX_i^{out,g}$ are listed and summarized in Table \ref{tab:covs}. Outcome-unit covariates $\vX_i^{out}$ include characteristics of the general population living in ZIP code $i$ (e.g., population, population density, percent Hispanic, high school graduation rate, median household income, poverty, occupied housing, migration rate (\% of residents who moved within 5 years), smoking rate), climatalogical factors (temperature and relative humidity), characteristics of the Medicare population living in the ZIP code (average age, percent female beneficiaries, percent white beneficiaries, and percent black beneficiaries), and general measure of power plant pollution in the area according to the total HyADS influence from all power plants on the ZIP code. Interventional-unit covariates $\vX_j^{int}$ include characteristics of the power plants from 2005 such as the total number of controls for oxides of nitrogen (NO$_x$) emissions, the percent of units with Selective (non) Catalytic Reduction systems (a particular technology for NO$_x$ control), total heat input, total operating time, the average percent of operating capacity, and whether the plant participated in Phase II of the ARP.  In the power plant setting, only $\vX_j^{int}$ of the key-associated power plant for outcome unit $i$ are regarded as potential upstream confounders, implying that $\{\vX^{int,z}_{j}\}_{j \in t_i^\top} = \vX^{z,int}_{\jstar}$ and $\{\vX^{int,g}_{j}\}_{j \in t_i^\top} = \vX^{g,int}_{\jstar}$, and the threat of confounding due to these power plant characteristics is regarded as low in comparison to the ZIP code characteristics in $\vX_i^{out}$. This also implies the absence of confounding due to overlapping interference sets, that is, after conditioning on $(\vX_i^{out, z}, \vX_i^{out, g}$), characteristics of other ZIP codes do not relate to hospitalization rates at ZIP code $i$.  

\begin{figure}
    \centering
    \includegraphics{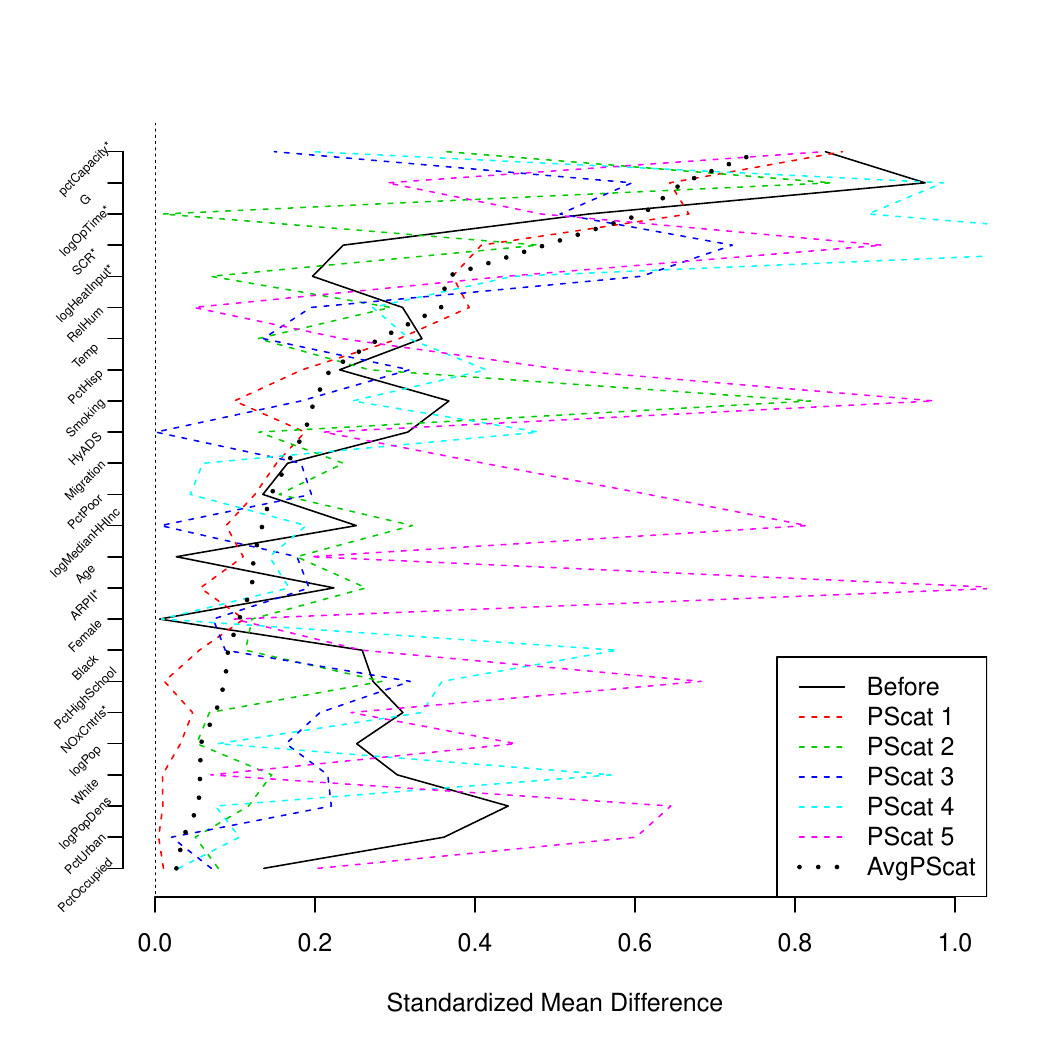}
    \caption{Balance Plot demonstrating covariate balance before any propensity score adjustment (Before), within each subclass of the individual propensity score (PScat), and averaged across all subclasses (AvgPScat) Variables marked with $*$ are individual power plant characteristics where the threat of confounding is considered less pronounced. Note that variable $G$ is a treatment, and is not a confounder.}
    \label{fig:balplot}
\end{figure}

The model for the key-associated propensity score, $\phi(z;x^{int,z}, x^{out,z})$, specifies $f^Z(z,\{\vX^{int,z}_{j}\}_{j \in t_i^\top}, \vX_i^{z,out};\gamma)$ as a logistic regression with main effect terms for each of $\vX^{int, z}_{\jstar}, \vX^{out,z}_i$, denoting the each of the power plant characteristics (of the key-associated plant) and each of the ZIP code characteristics listed in Table \ref{tab:covs}.  Estimates of this model are then used to first prune 626 ZIP codes for having estimates of $\hat{\phi}_i$ that did not overlap with the opposite treatment group, and then stratify each of the remaining $n=\numzips$ ZIP codes into one of $K=5$ strata based on the quintiles of the distribution of $\hat{\phi}_i$ among the ZIP codes with $Z_i=1$.  Figure \ref{fig:balplot} depicts the standardized mean difference in each covariate before the stratification, within each of the $K$ strata, and on average across all strata.  Note that, while the average standardized mean covariate difference across all strata is generally reduced relative to the unadjusted differences, serious imbalance remains, in particular within individual propensity score strata, motivating the use of further covariate adjustment in addition to adjustment for $\hat{\lambda}_i$, as will be described later.  Further note that the most extreme imbalances remain for characteristics of key-associated power plants for which the threat of confounding is judged to be minor compared to ZIP code characteristics that tend to exhibit better balance.

The model for the upwind propensity score, $\lambda(g; z, x^{int,g}, x^{out,g})$, specifies $f^G(g,z,\{\vX^{int,g}_{j}\}_{j \in t_i^\top}, \vX^{out,g}_i; \delta^k)$ as a normal regression model with linear main effect terms for only the ZIP code characteristics listed in Table \ref{tab:covs} ($\vX^{out,g}_i$).


\begin{table}[ht]
\caption{Covariate summary across levels of the key-associated treatment, $Z$.}\label{tab:covs}
\centering
\begin{tabular}{clrrr}
  \hline
 & & Mean Z=0 & Mean Z=1 \\   \hline
  \hline
&G & 0.49 & 0.86 \\ 
 & log(population) & 8.20 & 8.62 \\ 
 & \% Urban & 0.38 & 0.54 \\ 
 & \% Hispanic & 0.03 & 0.05 \\ 
 & \% High School Grad & 0.36 & 0.33 \\ 
 & log(MedianHouseholdIncome) & 10.52 & 10.62 \\ 
 & \% Poverty & 0.13 & 0.11 \\ 
 & \% Occupied Housing & 0.88 & 0.89 \\ 
 & Migration Rate & 0.41 & 0.44 \\ 
 ZIP Code & log$(\frac{pop}{mi^2})$ & 4.97 & 5.87 \\ 
Characteristics, $\vX^{out}$ & Smoking Rate & 0.26 & 0.25 \\ 
 & Temperature & 286.67 & 288.27 \\ 
 & Relative Humidity & 0.01 & 0.01 \\ 
  & Age & 74.95 & 74.99 \\ 
 & \% Female & 0.56 & 0.56 \\ 
 & \% White & 0.90 & 0.84 \\ 
 & \% Black & 0.08 & 0.13 \\ 
 & Total HyADS & 3.34 & 3.95 \\ 
 \hline
 & Total NO$_x$ Controls & 4.38 & 5.49 \\ 
 & log(HeatInput) & 14.95 & 14.77 \\ 
 Power Plant & log(OperatingTime) & 7.78 & 7.56 \\ 
 Characteristics, $\vX^{int}$ & \% Operating Capacity & 0.61 & 0.75 \\ 
 & \% of Selective (non) Catalytic Reduction & 0.18 & 0.27 \\ 
 & ARP Phase II & 0.70 & 0.59 \\ 
   \hline
\end{tabular}
\end{table}

The outcome model for estimating $E[Y_i(Z_i=z, G_i=g) | i \in K_k]$ specifies $f_k^y(z,g,\hat{\lambda};\theta_k)$ as a Poisson regression of the form:
\begin{align*}
log\biggl(\frac{Y_i(z,g)}{Beneficiaries_i}\biggr)=&\beta_0+\beta_z z+\beta_{g} g+ \beta_{\lambda} \hat{\lambda}_i+ \beta_{\lambda g}\hat{\lambda}_ig + \beta^\top_X\vX^{out}_i
\end{align*}
where $Beneficiaries_i$ is the number of Medicare fee-for-service person-years at risk for ZIP code $i$ in 2005 and $\vX^{out}_i$ contains all the ZIP code characteristics in Table \ref{tab:covs} to adjust for residual confounding not captured by the key-associated propensity score subclassification.

\subsection{Power Plant Analysis Results for IHD Hospitalization}
\begin{figure}
    \centering
    \includegraphics{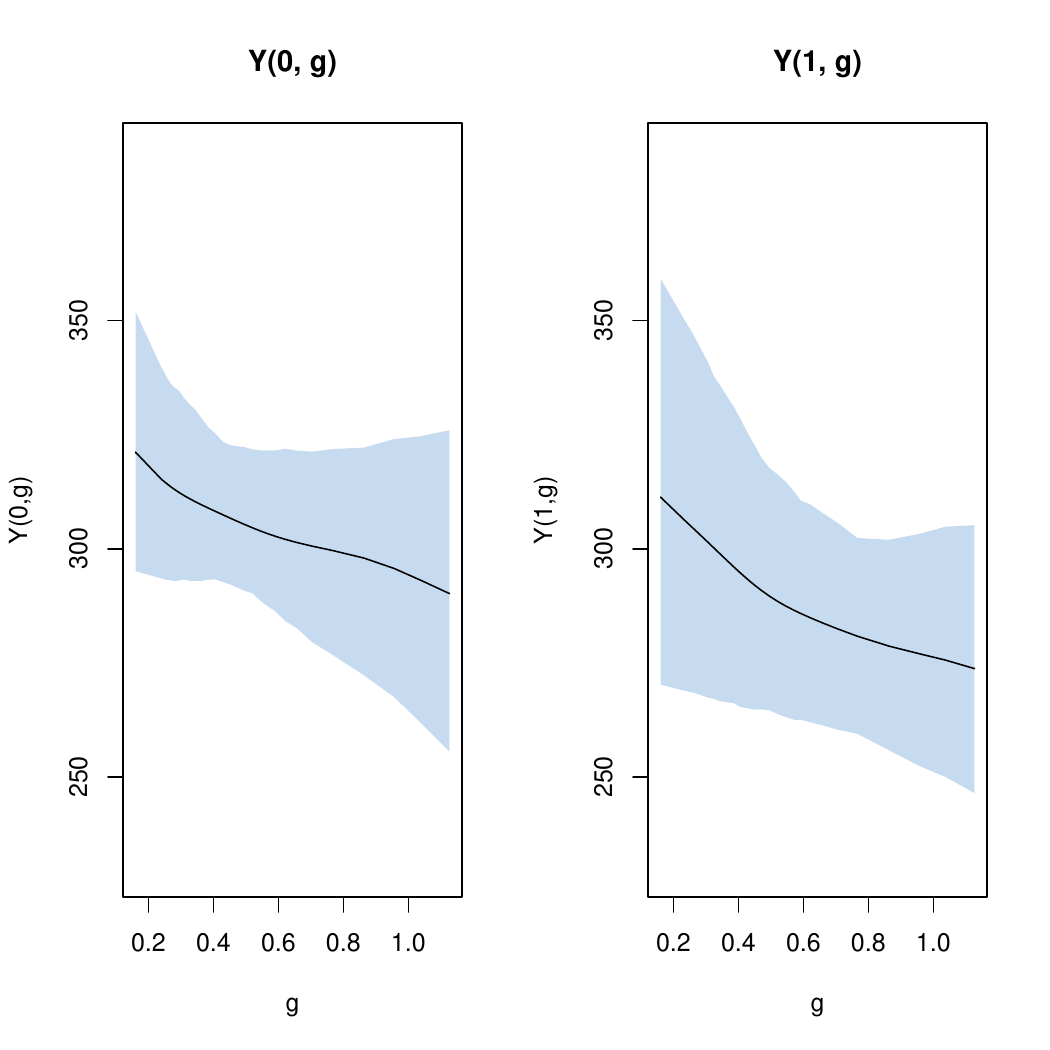}
    \caption{Estimated dose-response curves, where $Y(z,g)$ represents the Medicare IHD hospitalization rate per 10,000 person-years}
    \label{fig:drcurves}
\end{figure}

An analysis with 500 bootstrap samples for standard error estimation estimates $\hat{\tau} = -22.82 (-38.73, 14.42)$, indicating that, on average, having a scrubber installed on a ZIP code's most influential power plant causes a reduction of approximately 23 hospitalizations per 10,000 person-years, although this estimate cannot be conclusively distinguished from zero. The upwind treatment effect is estimated to be $\hat{\Delta}(0) = -14.37 (-36.83, 5.70)$ among ZIP codes for which the most influential power plant is without scrubber, and $\hat{\Delta}(1) = -17.69 (-39.89, 2.46)$ among the ZIP codes for which the most influential power plant has a scrubber, both suggestive of reduction in hospitalizations due to upwind scrubbers, but not conclusively different from zero.  Figure \ref{fig:drcurves} shows the estimated dose-response curves of $Y(z, g)$ against $g\in [g^{min},1]$ for $z=0,1$, indicating a clear trend that more upwind scrubbers is associated with lower IHD hospitalization rates.

Note that the above causal estimates pertain only to the \numzips retained in the analysis (see Figure \ref{fig:locations_map}), representing ZIP codes that experience substantial power plant pollution and were not too (un)likely to have a scrubbed key-associated power plant (no power plants are omitted from the analysis).  To the extent that scrubbers impact air quality and/or health in ZIP codes omitted from the analysis, these impacts are not captured by this analysis.

\subsection{Approximate Validation: Analysis of Ambient \PMTwo}
\begin{figure}
    \centering
    \includegraphics{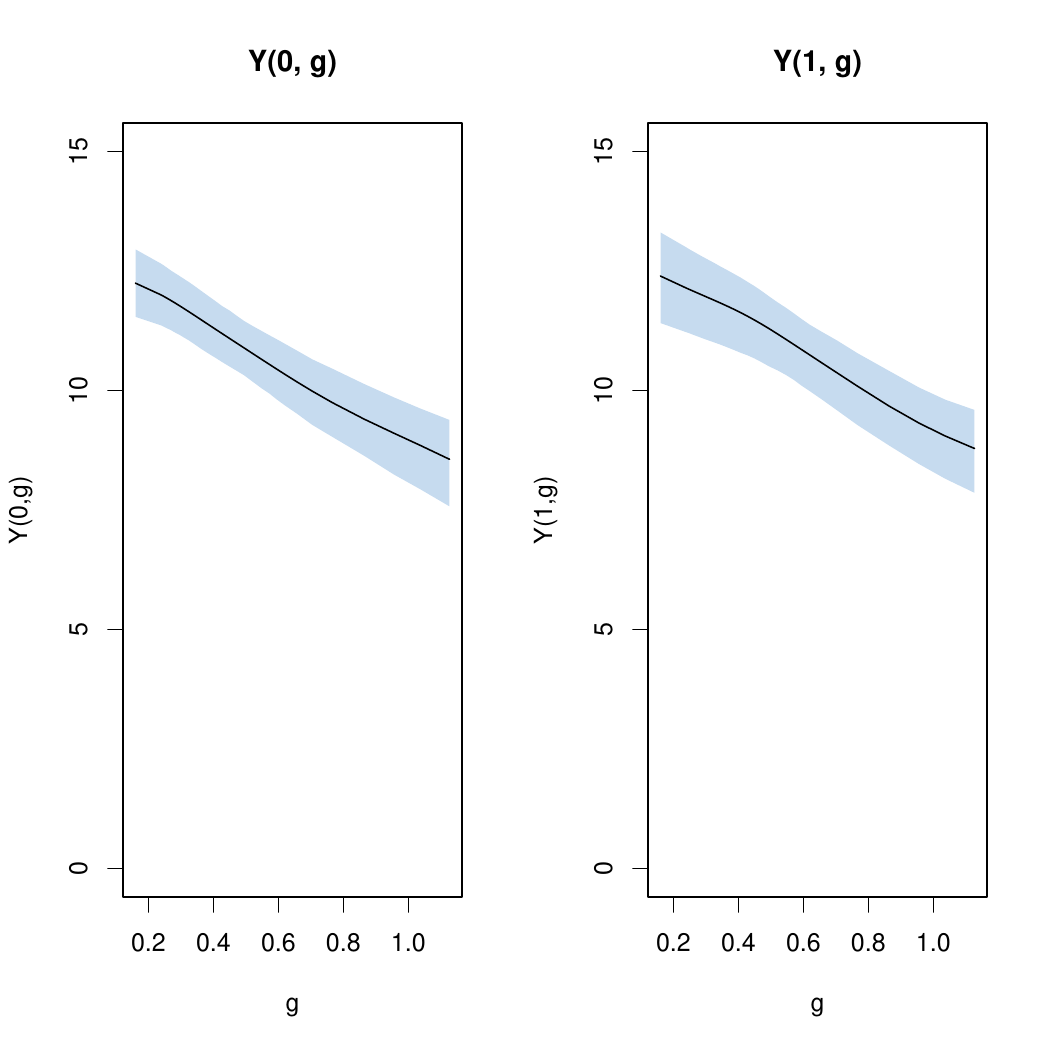}
    \caption{Estimated dose-response curves for the analysis where $Y(z,g)$ represents $\frac{\mu g}{m^3}$ of ambient \PMTwo{}.}
    \label{fig:drcurves_pm}
\end{figure}

For reference, we also conduct an analysis using the same model specification as described above, but with ambient \PMTwo as the outcome and the outcome model within subclasses of key-associated propensity scores specified as a normal regression with mean expression analogous to the Poisson regression in Section \ref{sec:model_spec}. This analysis can be viewed as a rough validation of some of the modeling infrastructure and assumptions about confounding entailed in the IHD analysis, as the link between power plant pollution and ambient \PMTwo is reasonably well understood, with a clear expectation that scrubbers reduce ambient \PMTwo and the regional nature of power plant pollution suggesting that collections of upwind scrubbers should dictate ambient \PMTwo more so than a scrubber at any single (key-associated) plant.  This analysis yields estimates of $\hat{\tau} = -0.37 (-0.32, 0.84)$; $\hat{\Delta}(0) = -1.34 (-1.75, -0.99)$; $\hat{\Delta}(1) = -1.16 (-1.63, -0.80)$, with depiction of the dose-response curves between $g$ and $Y(z, g)$ for $z=0,1$ in Figure \ref{fig:drcurves_pm}.  Thus, the analysis suggests that having a scrubber on the key-associated power plant may reduce ambient \PMTwo (measured in $\frac{\mu g}{m^3}$), with a clear signal that larger rates of upwind scrubbers lead to lower ambient \PMTwo.  For reference, the federal ambient air quality standard for annual average \PMTwo is 12 $\frac{\mu g}{m^3}$, and a reduction of 0.4 $\frac{\mu g}{m^3}$ attributable to any single source is substantial.  Estimates that are consistent with expectations given extant knowledge of how power plants contribute to ambient \PMTwo provides some degree of confidence in the modeling approach and the inclusion of important confounders.

\section{Discussion}
We have offered new estimands and a corresponding estimation strategy for causal effects on a bipartite network.  The investigation was specifically motivated by a problem in air pollution regulatory policy where scrubbers installed at coal-fired power plants are investigated for their effectiveness for reducing Medicare IHD hospitalizations, but hold relevance for other types of interventions where the units on which the treatments are defined (here, coal-fired power plants) are distinct from the units on which outcomes are relevant (here, ZIP codes), and the complex nature of relationships between these distinct types of unit lead to interference (market experiments are an emerging common example \citep{pouget-abadie_variance_2019, doudchenko_causal_2020, harshaw_design_2021}). 

While the GPS estimation strategy leverages the properties proposed and proven in  \cite{forastiere_identification_2021}, deploying this type of methodology in the setting of a bipartite network where interference arises due to a complex physical process entailed several distinctions and extensions over the more typical analysis of a social network.  First, the bipartite nature of the problem complicates standard notions of ``direct,'' ``indirect,'' and ``spillover,'' since there is no immediate correspondence governing which interventions apply ``directly'' or ``indirectly'' to an outcome unit.  The approach here provided a set of estimands that rely on specification of a key-associated interventional unit for each outcome unit, representing a subset of bipartite estimands proposed in \cite{zigler_bipartite_2021} that hold particular relevance in the power plant investigation.  Relying on the notion of the key-associated and upwind treatments in the bipartite setting introduced further differentiation with the work in \cite{forastiere_identification_2021}.  For example, this formulation presented the possibility of a joint assignment mechanism (for $Z$ and $G$) that naturally corresponds to independent variability in the two treatment components, even for a fixed network.  This is an important distinction with similar approaches in the setting with one level of observational unit, where ``neighborhood'' treatments analogous to $G$ would be deterministically governed by the structure of the network, $T$, and the allocation of treatments to individual units, $Z$.  Thus, in the bipartite setting, the indirect effect proposed in Section \ref{sec:estimands} might correspond more naturally to an actual and implementable intervention that changes allocations to some interventional units (and, as a consequence, the level of $G$) without changing $Z$. It is important to note that this formulation relies on both an appropriate definition of ``key-associated'' and the assumption that the key-associated unit, however defined, is appropriately designated.  This work defined ``key-assocaited'' based on a characteristic of the adjacency matrix, $T$, but natural alternatives could be defined based on immutable characteristics such as geographic distance, which may be important when regarding $T$ as time-varying or uncertain.  Sensitivity analyses to the ``key-associated'' designation, including those to accommodate the possibility that the true ``key-associated'' unit may not appear in the data, are an interesting topic of future work. 

The type of interference considered here is due to complex exposure patterns, which is an important distinction with most existing work where interference arises from unit-to-unit outcome dependencies.  This was framed as a problem of interference on a weighted, directed network, expanding common notions of network dependency and adjacency that arise in settings where interference arises due to outcome dependence among one level of observational unit. Issues related to spatial correlation, homophily, and confounding all take on somewhat different meanings than those that have become routine in studies of adjacency networks.  More broadly, the particulars of the power plant investigation anchored the deployment of causal inference methods to confront interference with data that are explicitly spatially-indexed. Despite early progress in \cite{verbitsky-savitz_causal_2012} and recent advances in \cite{giffin_generalized_2020, aronow_design-based_2020, zirkle_addressing_2021}, the literature on explicit and potential-outcomes based methods for spatial interference remains sparse.

In addition to the contributions to statistical methodology, this work represents the first (to our knowledge) rigorous application of methods for causal inference with interference in air pollution that attempts to reflect the complex atmospheric processes underlying the interference.  A previous analysis in \cite{zigler_bipartite_2021} relied on ad-hoc clustering, which was known to be a gross simplification of interference in this context. The combination of the data sources described in Section \ref{sec:background} with the novel reduced-complexity atmospheric model (HyADS) to characterize the structure of interference represents an important advance in environmental data science at the intersection of statistics and atmospheric science.  What's more, the formalization of potential outcomes to focus on causal estimands indexed by discrete interventions at power plants (i.e., the installation or not of a scrubber) and acknowledge interference is an important advance over previous epidemiological investigations that deploy HyADS to characterize locations' cumulative exposure to a set of power plants, without maintaining the explicit distinction between which of a set of power plants did or did not take a particular action \citep{henneman_accountability_2019}.  Results from an analysis such as this should be interpreted alongside those of other (non-statistical) pollution modeling efforts which more directly model the impacts from individual power plants and attribute ``per unit'' impacts on, say, ambient \PMTwo.  For example, an average causal effect such as $\tau$ cannot be applied to all \numplants power plants to estimate the air quality that would occur if every power plant installed a scrubber, as it represents an average over units and over the observed distribution of $G$, and air quality impacts of actions taken across many point sources are likely non-additive. 

Despite the advances in methods for causal inference and analysis of air quality policy, there are important limitations to this work.  First is the interventional-unit bootstrap method used for inference, which was shown to be conservative in the simulation study, relies on the ``egocentric'' network sampling mechanism where interventional units are individually resampled, but network-derived quantities such as the key-associated and upwind treatment are regarded as fixed characteristics of the units.  This perspective when applied to the power plant investigation also relied on a model-based perspective for inference where potential outcomes are regarded as random variables with values drawn from a specified model. This motivates exploration of alternative paradigms for inference, possibly including a superpopulation perspective applied to a set of fixed points, which would have points of contact with the spatial statistics literature \cite{cressie_statistics_2015}.  Second, the estimation strategy of first stratifying outcome units on the key-associated propensity score and then fitting parametric models within strata for the upwind propensity score and the potential outcomes represents one reasonable approach, but the threat of confounding remained particularly pronounced in the lack of covariate balance for many ZIP code features within subclasses of the key-associated propensity score, with alternative strategies for propensity score adjustment producing similarly unsatisfactory covariate balance. The direct adjustment for ZIP code level covariates in the dose-response models is expected to account for residual confounding, but should nonetheless be interpreted with caution due to the reliance on parametric modeling assumptions.  Other, more flexible approaches to adjust for confounding deserve further exploration  particularly those that might explicitly account for spatial correlation or the possibility that power plants owned by the same corporation make dependent choices about which plants receive scrubbers. The analysis presented here adjusted for many features of power plants, population demographics, and weather, but a key source of potential unmeasured confounding is \PMTwo that derives from sources {\it other} than coal-fired power plants, for example, from vehicular traffic.  Plausibility of the ignorability assumption relates to the presumption that other sources of \PMTwo that are systematically related to scrubber installation and IHD hospitalizations are likely related to measured ZIP code metrics such as population density. Considerations such as this highlight that confounding in the power plant investigation is a consequence of {\it both} operator decisions about scrubber installation {\it and} idiosyncrasies relating to pollution transport and the distribution of populations across space.  This motivates the joint GPS approach that simply attempts to balance outcome-unit characteristics across levels of ($Z_i, G_i$), although the threat of unmeasured confounding remains. Finally, the analysis pursued here condenses the clearly time-varying nature of the interventions and network structure to a single annual summary.  Scrubbers are installed on additional power plants over time, and the underlying dynamics of pollution transport vary continuously, with particularly pronounced seasonal variation within a year as well as decades-long variation owing to other atmospheric changes. Thus, while we construe $T$ to be fixed at its annual summary, our analysis does not reflect any possible uncertainty in $T$ (owing to HyADS modeling errors) and, in reality, the structure of the network interference evolves over time. Further development of time-varying treatments on time-varying interference networks is an important area of future work that holds particular relevance to the analysis of power plant regulations.  


\section*{Acknowledgments}
This work was supported by research funding from NIH R01ES026217, US EPA 83587201, and Dipartimenti Eccellenti 2018-2022 Italian Ministerial Funds. Its contents are solely the responsibility of the grantee and do not necessarily represent the official views of the USEPA. Further, USEPA does not endorse the purchase of any commercial products or services mentioned in the publication.

\section*{References}
\bibliography{PowerPlantsR01}

\end{document}